\documentclass{aa}

\usepackage{graphicx}
\usepackage{txfonts}
\usepackage{amsmath, amssymb}
\usepackage{mathtools}
\usepackage{commath}
\usepackage{natbib}
\usepackage{siunitx}
\usepackage{array}
\usepackage[]{hyperref}

\makeatletter
\renewcommand*\aa@pageof{, page \thepage{} of \pageref*{LastPage}}
\makeatother

\usepackage{color}
\usepackage[dvipsnames]{xcolor}

\DeclareSIUnit[]\solarradius{R_\odot}
\DeclareSIUnit[]\year{yr}

\defcitealias{Loeptien2018}{LGBS18}
\defcitealias{Liang2019}{LGBD19}

\begin{document}

\title{Exploring the latitude and depth dependence of solar Rossby waves using ring-diagram analysis}

\author{
B.~Proxauf\inst{1} \and 
L.~Gizon\inst{1,2,3} \and 
B.~L{\"o}ptien\inst{1} \and 
J.~Schou\inst{1} \and 
A.~C.~Birch\inst{1} \and 
R.~S.~Bogart\inst{4}
}

\institute{
Max-Planck-Institut f{\"u}r Sonnensystemforschung, Justus-von-Liebig-Weg 3, 37077 G{\"o}ttingen, Germany \\
\email{proxauf@mps.mpg.de} \and 
Institut f{\"u}r Astrophysik, Georg-August-Universit{\"a}t G{\"o}ttingen, Friedrich-Hund-Platz 1, 37077 G{\"o}ttingen, Germany \and 
Center for Space Science, NYUAD Institute, New York University Abu Dhabi, Abu Dhabi, UAE \and 
W.~W. Hansen Experimental Physics Laboratory, Stanford University, Stanford, CA 94305, USA
}

\date{Received TBD; accepted TBD}

% \abstract{}{}{}{}{}
% 5 {} token are mandatory

\abstract
% context heading (optional), leave it empty if necessary
{Global-scale equatorial Rossby waves have recently been unambiguously identified on the Sun. Like solar acoustic modes, Rossby waves are probes of the solar interior.}
{We study the latitude and depth dependence of the Rossby wave eigenfunctions.}
{By applying helioseismic ring-diagram analysis and granulation tracking to observations by HMI aboard SDO, we computed maps of the radial vorticity of flows in the upper solar convection zone (down to depths of more than $\SI{16}{\mega\metre}$). The horizontal sampling of the ring-diagram maps is approximately $\SI{90}{\mega\metre}$ ($\sim \SI{7.5}{\degree}$) and the temporal sampling is roughly $27$~hours. We used a Fourier transform in longitude to separate the different azimuthal orders $m$ in the range $3 \leq m \leq 15$. At each $m$ we obtained the phase and amplitude of the Rossby waves as functions of depth using the helioseismic data. At each $m$ we also measured the latitude dependence of the eigenfunctions by calculating the covariance between the equator and other latitudes.}
{We conducted a study of the horizontal and radial dependences of the radial vorticity eigenfunctions. The horizontal eigenfunctions are complex. As observed previously, the real part peaks at the equator and switches sign near $\pm\SI{30}{\degree}$, thus the eigenfunctions show significant non-sectoral contributions. The imaginary part is smaller than the real part. The phase of the radial eigenfunctions varies by only $\pm\SI{5}{\degree}$ over the top $\SI{15}{\mega\metre}$. The amplitude of the radial eigenfunctions decreases by about $\SI{10}{\percent}$ from the surface down to $\SI{8}{\mega\metre}$ (the region in which ring-diagram analysis is most reliable, as seen by comparing with the rotation rate measured by global-mode seismology).}
{The radial dependence of the radial vorticity eigenfunctions deduced from ring-diagram analysis is consistent with a power law down to $\SI{8}{\mega\metre}$ and is unreliable at larger depths. However, the observations provide only weak constraints on the power-law exponents. For the real part, the latitude dependence of the eigenfunctions is consistent with previous work (using granulation tracking). The imaginary part is smaller than the real part but significantly nonzero.}


\titlerunning{Latitude and depth dependence of solar Rossby waves}
\maketitle

\section{Introduction}
\label{sect_introduction}

Recently, \citet[][hereafter \citetalias{Loeptien2018}]{Loeptien2018} discovered global-scale Rossby waves in maps of flows on the surface of the Sun. These waves are waves of radial vorticity that may exist in any rotating fluid body. Even though Rossby waves were predicted to exist in stars more than $40$~years ago \citep{Papaloizou1978, Saio1982}, solar Rossby waves were difficult to detect because of their small amplitudes ($\sim \SI{1}{\metre\per\second}$) and long periods of several months. Solar Rossby waves contain almost as much vorticity as large-scale solar convection. The dispersion relation of solar Rossby waves is close to the standard relation for sectoral modes, $\omega = -2\Omega/(m + 1)$, where $\Omega$ is the rotation rate of a rigidly rotating star and $m$ is the azimuthal order \citep{Saio1982}. Rossby waves have a retrograde phase speed and a prograde group speed. In \citetalias{Loeptien2018}, the authors also measured the horizontal eigenfunctions, which peak at the equator.

The detection of solar Rossby waves was confirmed by \citet[][hereafter \citetalias{Liang2019}]{Liang2019} with time-distance helioseismology \citep{Duvall1993} using data covering more than $20$~years, obtained from the Solar and Heliospheric Observatory (SOHO) and from the Solar Dynamics Observatory \citep[SDO;][]{Pesnell2012}. \citet{Alshehhi2019}, in an effort to speed up ring-diagram analysis \citep[RDA;][]{Hill1988} via machine learning, also saw global-scale Rossby waves. \citet{Hanasoge2019} and \citet{Mandal2019} provide another recent Rossby wave confirmation using a different technique of helioseismology known as normal-mode coupling \citep{Woodard1989, Hanasoge2017}.

Knowledge about the latitude dependence of Rossby wave eigenfunctions is incomplete, as \citetalias{Loeptien2018} studied only their real parts. In a differentially rotating star, the horizontal eigenfunctions are not necessarily spherical harmonics (and may not even separate in latitude and depth). Also, little is known observationally about the depth dependence of the Rossby waves. It would be well worth distinguishing between the few existing theoretical models of the depth dependence \citep{Provost1981, Smeyers1981, Saio1982, Wolff1986}.

In this paper, we explore the latitude dependence of the eigenfunctions, as well as the phase and amplitude of solar Rossby waves as functions of depth from the surface down to more than $\SI{16}{\mega\metre}$ using helioseismology. We use observations from the Helioseismic and Magnetic Imager \citep[HMI;][]{Schou2012} on board SDO, processed with RDA. From these we attempt to measure the eigenfunctions of the Rossby waves in the solar interior. For comparison near the surface, we also use data from local correlation tracking of granulation \citep[LCT;][]{November1988}.

\section{Data and methods}
\label{sect_data_methods}

We used maps of the horizontal velocity, derived from two different techniques applied to SDO/HMI observations. The first dataset consists of LCT (granulation tracking) flow maps at the surface \citep{Loeptien2017} and covers almost six years from May 20, 2010 to March 30, 2016. The second dataset comprises RDA flow maps from the HMI ring-diagram pipeline (\citealt{Bogart2011a, Bogart2011b}; see also \citealt{Bogart2015}). For comparisons with LCT, we took a period as close to the LCT period as possible, i.e., May 19, 2010 to March 31, 2016, while for all other results we used a longer period of more than seven years from May 19, 2010 to December 29, 2017; this corresponds to 102 Carrington rotations (CRs), i.e., CR $2097$ - $2198$.

\subsection{Overview of LCT data}

The LCT flow maps are obtained from and processed as described in \citet{Loeptien2017}. They are created by applying the Fourier LCT code \citep[FLCT;][]{Welsch2004, Fisher2008} to track the solar granulation in pairs of consecutive HMI intensity images. The image pairs are separated by $30$~minutes. Several known systematic effects such as the shrinking-Sun effect \citep{Lisle2004, Loeptien2016} and effects related to the SDO orbit are present in the LCT maps. Therefore the maps are decomposed into Zernike polynomials, a basis of 2D orthogonal functions on the unit disk, and the time series of the coefficient amplitudes for the lowest few Zernike polynomials are filtered to remove frequencies of one day and one year (associated with the SDO orbit) as well as all harmonics up to the Nyquist frequency. The zero frequency is also removed. The filtered maps are then tracked at the sidereal Carrington rate and remapped onto an equi-spaced longitude-latitude grid with a step size of $\SI{0.4}{\degree}$ in both directions.

\subsection{Overview of ring-diagram data}
\label{sect_data_methods_rda_overview}

The ring-diagram pipeline \citep{Bogart2011a, Bogart2011b} takes HMI Dopplergrams as input and remaps them onto tiles spanning $182 \times \SI{182}{\mega\metre}$ (i.e., $\SI{15}{\degree}$ each in latitude $\lambda$ and longitude $\varphi$ at the equator). The tiles overlap each other by roughly $\SI{50}{\percent}$ in each direction such that the tile borders fall onto the centers of adjacent tiles. Both the latitude and longitude sampling are half the tile size. The latitude grid is linear and includes the equator, while the longitude grid is also linear, but is latitude-dependent. Each tile is tracked for $1728$~minutes ($28.8$~hours) at the sidereal Carrington rate. The temporal grid spacing is, on average, $1/24$ of the synodic Carrington rotation period of $27.2753$~days.

In the pipeline, for each tile a 3D local power spectrum is computed from the tracked Dopplergrams. The velocity fit parameters $U_{x,\,n \ell}$ (prograde) and $U_{y,\,n \ell}$ (northward) are extracted via a ring-fit algorithm \citep{Haber2000} for different solar oscillation modes, which are indexed by their radial order $n$ and angular degree $\ell$. The flow velocities $u_x$ and $u_y$ are inferred for various target depths via a 1D optimally localized averages (OLA) inversion. The inversion results for the six-parameter fits of the $\SI{15}{\degree}$ tiles sample a range of target depths from $\SI{0.97}{R_\odot}$ to $\SI{1}{R_\odot}$ (step size $\SI{0.001}{R_\odot}$), corresponding to a nonlinear grid of measurement depths (median of the ring-diagram averaging kernels) from $\SI{0.976}{R_\odot}$ to $\SI{1}{R_\odot}$. In this paper, the term depth always refers to measurement depth and not to target depth.

The inversion results are stored in the Joint Science Operations Center (JSOC) data series hmi.V\_rdvflows\_fd15\_frame. However, up to inversion module rdvinv v.0.91, the inversion results depended on the input tile processing order due to an array initialization bug. This caused significantly lower velocity uncertainties for tiles near latitude $\SI{7.5}{\degree}$ and Stonyhurst longitude $\SI{37.5}{\degree}$, even when averaged over seven years, but also slightly affected the velocities. At the same disk locations the bug caused a correlation of $u_x$ with the $B_0$ angle. Since rdvinv v.0.92 is officially only applied since March 2018, we re-inverted the entire dataset ourselves for the analysis shown in this work.

Apart from the array initialization bug, we found several other issues with the default HMI ring-diagram pipeline that have not yet been solved. Among these are under-regularization in the inversion for some individual tiles, leading to relatively narrow averaging kernels and anomalously high noise. Finally, the number of ring fits used for the inversion depends strongly on disk position. This may lead to systematic effects and additional noise.

The ring-diagram velocities $u_x$ reported at a certain measurement depth $r$ at the equator for an angular rotation rate $\Omega(r)$ are equal to $\Omega(r) R_\odot$ instead of the local velocity $\Omega(r) r$. Since we are interested in the latter, we multiplied $u_x$ by $r/R_\odot$. By analogy, we also applied this factor to $u_y$ and to all other latitudes. Additionally, the inversion does not account for the quantity $\beta_{n \ell}$, defined, for example, in Eq.~3.357 of \citet{Aerts2010}. The quantity $\beta_{n \ell}$ is related to the effect of the Coriolis force on the mode frequency splitting. For uniform rotation in particular, at fixed $m$, $\beta_{n \ell}$ completely describes the effect of the rotation on the mode frequency splitting. Both issues are described in more detail in App.~\ref{app_rda_inversion_issues}.

\subsection{Post-processing of ring-diagram data}

The ring-diagram data are organized in CRs, which undergo several processing steps, including the removal of systematic effects, an interpolation in longitude, an interpolation in time, and the removal of limb data.

Several systematic effects are present in the ring-diagram velocities, such as center-to-limb effects that depend on the disk position of the tile \citep{Baldner2012, Zhao2012}. There are time-independent effects and systematics with a one-year period, which are probably related to the $B_0$ angle. To remove the systematics, we fit the time series at each position on the disk (in Stonyhurst coordinates) with sinusoids
\begin{equation}
\begin{aligned}
& u_x(t) = a_x \sin(2\pi t/(\SI{1}{yr})) + b_x \cos(2\pi t/(\SI{1}{yr})) + c_x, \\
& u_y(t) = a_y \sin(2\pi t/(\SI{1}{yr})) + b_y \cos(2\pi t/(\SI{1}{yr})) + c_y,
\end{aligned}
\end{equation}
and subtract the fits from the flow velocities. We used all available CRs to determine the fit parameters.

Because of the specific tile coordinate selection used by the ring-diagram pipeline \citep{Bogart2011a}, which seeks to optimally cover the visible disk, tile centers at different latitudes have Stonyhurst longitudes that are offset by multiples of $\SI{2.5}{\degree}$ from each other. To obtain a latitude-independent longitude grid, we interpolated the flow maps in Stonyhurst longitude using splines (App.~\ref{app_rda_processing_steps}).

We also interpolated the ring-diagram flows in time similarly with splines to fill missing time steps due to instrumental issues (only $12$ out of $2448$ time steps), which cause a too low observational duty cycle ($\leq \SI{70}{\percent}$). We interpolated the data in the Carrington reference frame so as to use always roughly the same physical locations on the Sun. This mixes different systematics, which are primarily dependent on disk position, but we should already have removed the dominant contributions at this stage. We interpolated every missing time step from roughly the same number of data points (all available time steps within the corresponding disk passage) using splines (App.~\ref{app_rda_processing_steps}).

The output uncertainties from the ring-diagram pipeline increase strongly toward the limb, in particular beyond an angular great-circle distance of roughly $\SI{65}{\degree}$ to the crossing of the central meridian with the equator ($\lambda = \SI{0}{\degree}, \varphi = \SI{0}{\degree}$). We thus only used ring-diagram data within $\SI{65}{\degree}$ of ($\lambda = \SI{0}{\degree}, \varphi = \SI{0}{\degree}$).

\subsection{From velocity maps to power spectra of radial vorticity}

From this stage onward ring-diagram and LCT data are processed similarly. The processing steps include a shift to the equatorial rotation rate $\nu_{\text{eq}} = \Omega_{\text{eq}}/2\pi = \SI{453.1}{\nano\hertz}$, the subtraction of the longitude mean, the calculation of the radial vorticity, a spherical harmonic transform (SHT), and a Fourier transform of the SHT coefficient time series.

The flow maps are shifted from the tracking rate (sidereal Carrington rate) to the surface sidereal equatorial rotation rate of $\SI{453.1}{\nano\hertz}$, an average of zonal flows inferred from global-mode analysis of SDO/HMI observations \citep{Larson2018}. We shifted the LCT data in Fourier space via a time-dependent phase factor, applying the same convention for the Fourier transform as \citetalias{Loeptien2018}. The ring-diagram data are first apodized by a raised cosine in angular great-circle distance to the point ($\lambda = \SI{0}{\degree}, \varphi = \SI{0}{\degree}$) to suppress near-limb data and are shifted via spline-interpolation (App.~\ref{app_rda_processing_steps}).

We next subtracted the longitude mean from the data to remove any remaining large-scale flows. Differential rotation and meridional circulation should have already been subtracted in the RDA or LCT post-processing, but any possible longitude-independent flows still in the data are removed in this step.

Subsequently, we calculated the radial vorticity (via second-order central finite differences) as follows:
\begin{equation}
\begin{aligned}
\zeta(t,r,\lambda,\varphi) =& - \frac{1}{r \cos\lambda} \frac{\partial (u_x(t,r,\lambda,\varphi) \cos\lambda)}{\partial\lambda} \\
& + \frac{1}{r \cos\lambda} \frac{\partial u_y(t,r,\lambda,\varphi)}{\partial \varphi},
\end{aligned}
\end{equation}
where $r$ is the measurement depth. We decomposed the resulting maps into spherical harmonics and performed a temporal Fourier transform of the spherical harmonic coefficients. Last, we calculated the power and the phase (where the phase range is the half-open interval ($\SI{-180}{\degree},\SI{180}{\degree}$]). The sign convention is such that waves with positive $m$ and negative frequency $\nu$ have a retrograde phase speed.

If not stated otherwise, the terms power spectrum or Fourier transform used in this paper always refer to the power spectrum or Fourier transform of the radial vorticity. Similarly, we discuss eigenfunctions of radial vorticity. These eigenfunctions are not spherical harmonics, however \citepalias{Loeptien2018}. In particular, while the modes can be meaningfully indexed by $m$, the angular degree $\ell$ is not observable. Throughout the paper $\ell$ thus only refers to the projection of the Rossby wave modes onto the corresponding spherical harmonic and not to the Rossby wave eigenfunction itself. We also use the terms latitudinal and radial eigenfunctions, which assumes separability in the $r$ and $\lambda$ coordinates. This assumption is addressed in more detail in Sect.~\ref{sect_summary}.

\section{Results}
\label{sect_results}

\subsection{Radial vorticity maps}

\begin{figure}
\centering
\includegraphics[width=\hsize]{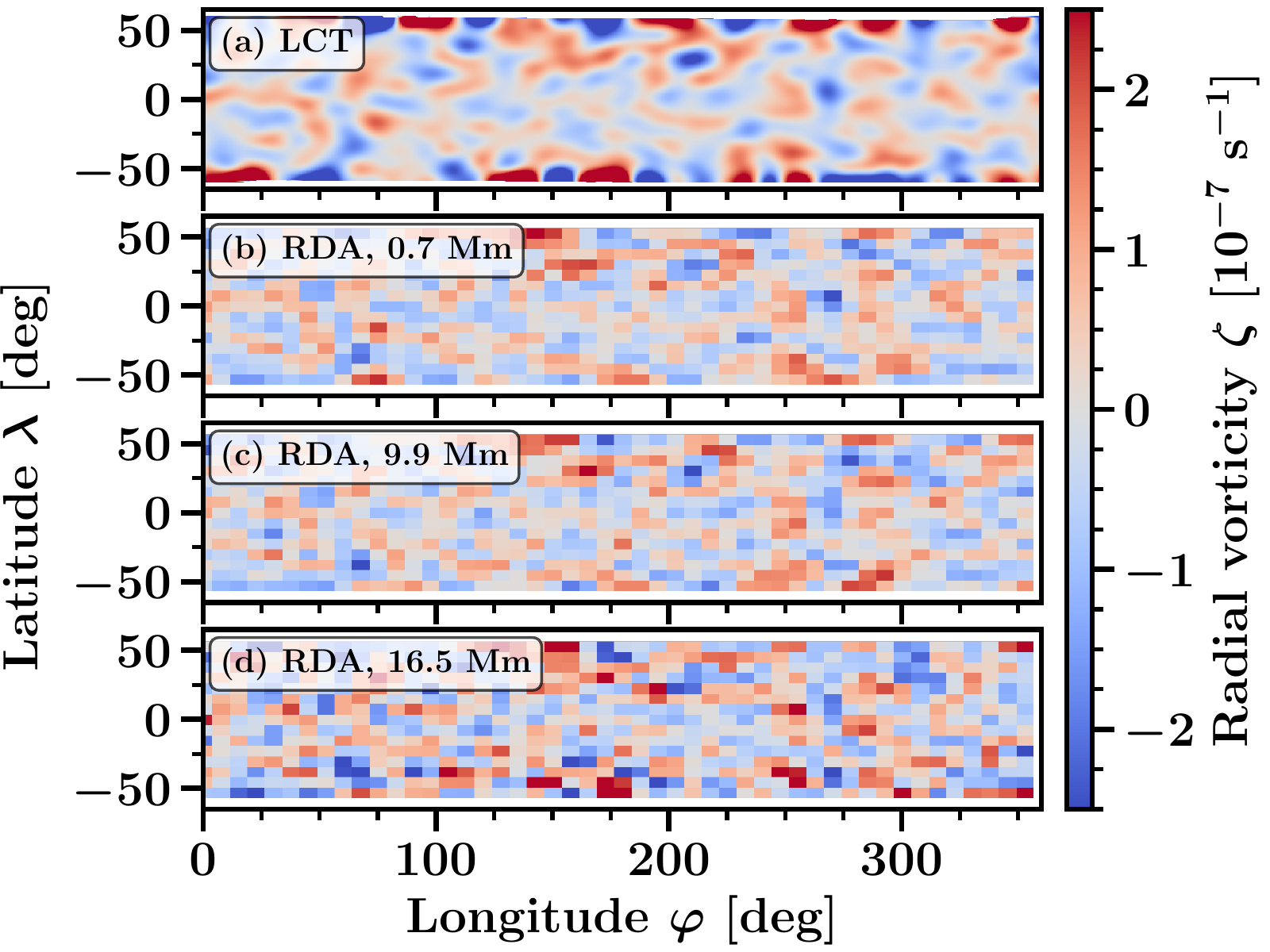}
\caption{
Radial vorticity maps from LCT at the surface and from RDA at depths $0.7$, $9.9$, and $\SI{16.5}{\mega\metre}$. The radial vorticity is averaged over one rotation (from May 20, 2010 to June 16, 2010). The LCT map is smoothed in latitude and longitude with a Gaussian filter ($\sigma = \SI{6}{\degree}$) to filter out small-scale convection.
}
\label{fig_vorticity_maps}
\end{figure}

Figure~\ref{fig_vorticity_maps} shows example vorticity maps from LCT surface flows and from RDA flows near the surface, at intermediate, and at large depths ($0.7$, $9.9$, and $\SI{16.5}{\mega\metre}$), averaged over the first rotation in the dataset (May 20, 2010 to June 16, 2010). The LCT data have a much better horizontal resolution than the ring-diagram data and thus pick up small-scale convective contributions. To be able to compare LCT with RDA, we thus smooth the LCT vorticity with 1D Gaussian filters of width $\sigma = \SI{6}{\degree}$ both in latitude and longitude.

We do not expect perfect agreement of the two methods because of their different sensitivities to horizontal scales and to different depths. Nonetheless, the LCT map shows similar features as the near-surface ($\SI{0.7}{\mega\metre}$) ring-diagram map. While large absolute radial vorticities are visible at high latitudes (beyond $\pm\SI{50}{\degree}$) in the LCT but not in the ring-diagram data, the vorticities near the equator agree. As a test, we interpolate the LCT data to the RDA grid using a 2D bicubic spline. The correlation coefficient between the interpolated LCT and the ring-diagram maps decreases with the latitude width of the strip of pixels considered and there is a steep decrease beyond $\pm\SI{45}{\degree}$. The correlation is 0.92 when including only equatorial pixels, 0.79 for pixels within $\pm\SI{45}{\degree}$, and 0.59 for all pixels, i.e., within $\pm\SI{52.5}{\degree}$. The noise increases strongly with depth (see lower panels of Fig.~\ref{fig_vorticity_maps}), but the main vorticity features are still visible.

\subsection{Power spectra of radial vorticity}
\label{sect_power_spectra_rossby}

\begin{figure}
\centering
\includegraphics[width=\hsize]{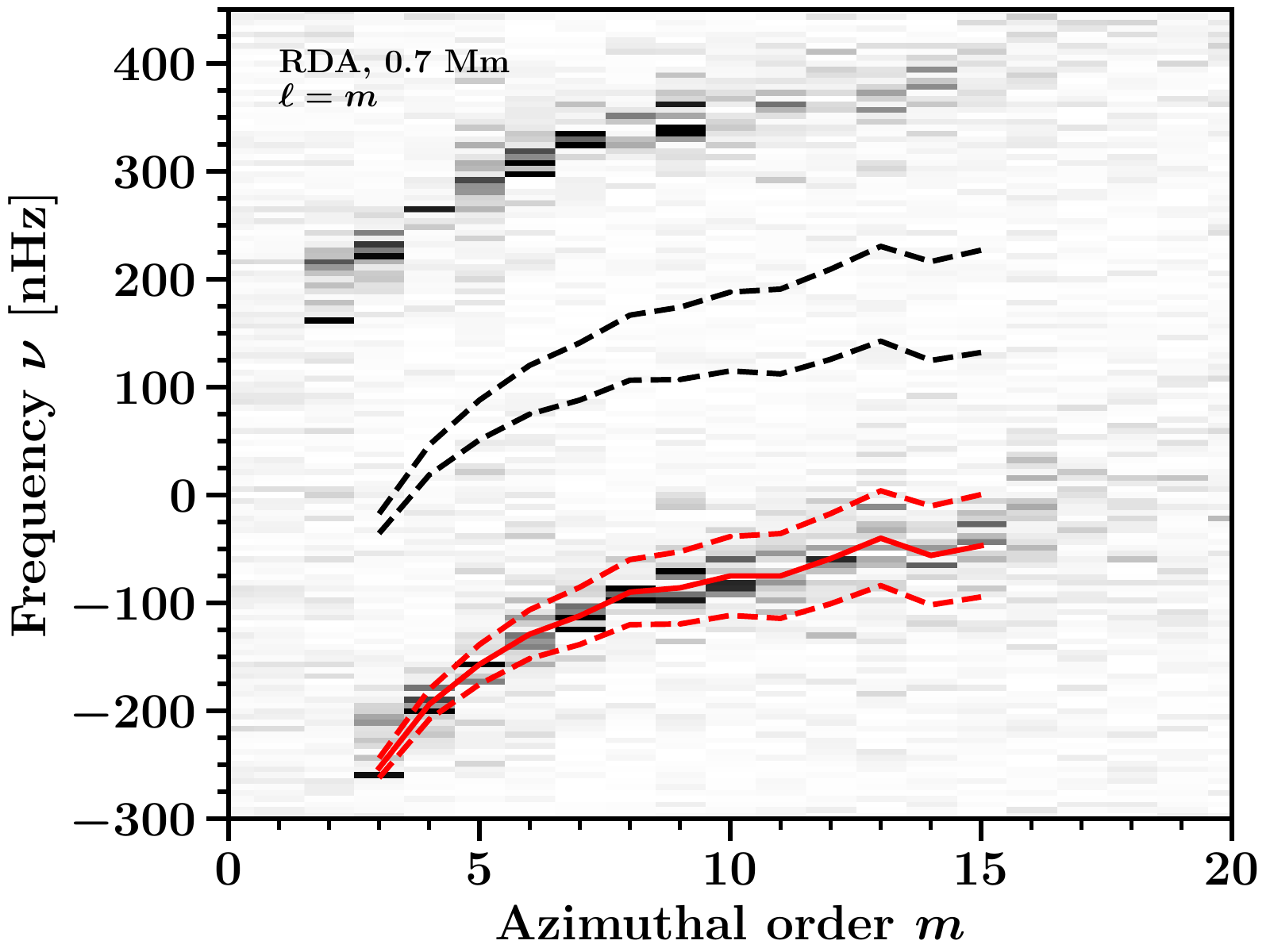}
\caption{
Sectoral power spectrum ($\ell = m$) of the radial vorticity for RDA data at depth $\SI{0.7}{\mega\metre}$. The solid red line indicates the Rossby wave frequencies from \citetalias{Liang2019} for $m = 3$ and from \citetalias{Loeptien2018} for $m \geq 4$. Frequency intervals for the Rossby wave region and the background region are indicated by the red and black dashed lines, respectively. The ridge of power at positive frequencies is due to the first side lobe of the window function. For better visibility of the Rossby waves at low $m$, the power is normalized at each $m$ by the frequency average over [$-300,100$]~nHz. The color scale is truncated at $\SI{50}{\percent}$ of the maximum value (black).
}
\label{fig_power_spectrum_2d}
\end{figure}

Figure~\ref{fig_power_spectrum_2d} shows the Rossby wave power of the $\ell = m$ component for the ring-diagram data near the surface ($\SI{0.7}{\mega\metre}$) versus frequency and azimuthal order $m$ (\citetalias{Loeptien2018} detected only the sectoral component of the Rossby waves). We divide the power, at each $m$, by the frequency average over [$-300,100$]~nHz near the surface ($\SI{0.7}{\mega\metre}$). The visible power ridge corresponds to the Rossby wave signal. The mode frequency increases with $m$ roughly according to the textbook dispersion relation for sectoral waves, $\omega = -2\Omega_{\text{eq}}/(m + 1)$, as seen earlier by \citetalias{Loeptien2018}.

Besides the Rossby wave signal there are other ridges, that, at fixed $\Delta m = m - m'$, are shifted from the Rossby waves by roughly $\Delta m \left(\nu_{\text{eq}} - 1/(\SI{1}{yr})\right)$, where $\nu_{\text{eq}} - 1/(\SI{1}{yr}) \sim \SI{421.4}{\nano\hertz}$. The main contribution to these side lobes comes from a temporal window function, which is introduced by solar rotation and not by time gaps; the time coverage is very good (see Sect.~\ref{sect_data_methods}). This leads to side lobes of wave power from modes at $m'$ to modes at $m$. We only show the side lobes for $\Delta m = +1$, but we typically observe the side lobes above the noise between $\Delta m = -2$ and $\Delta m = +3$. In \citetalias{Liang2019}, the authors use $21$~years of time-distance data from a combined sample of observations from the Michelson Doppler Imager (MDI) on board SOHO and from SDO/HMI and they discuss the window function in detail.

\begin{figure*}
\centering
\includegraphics[width=\hsize]{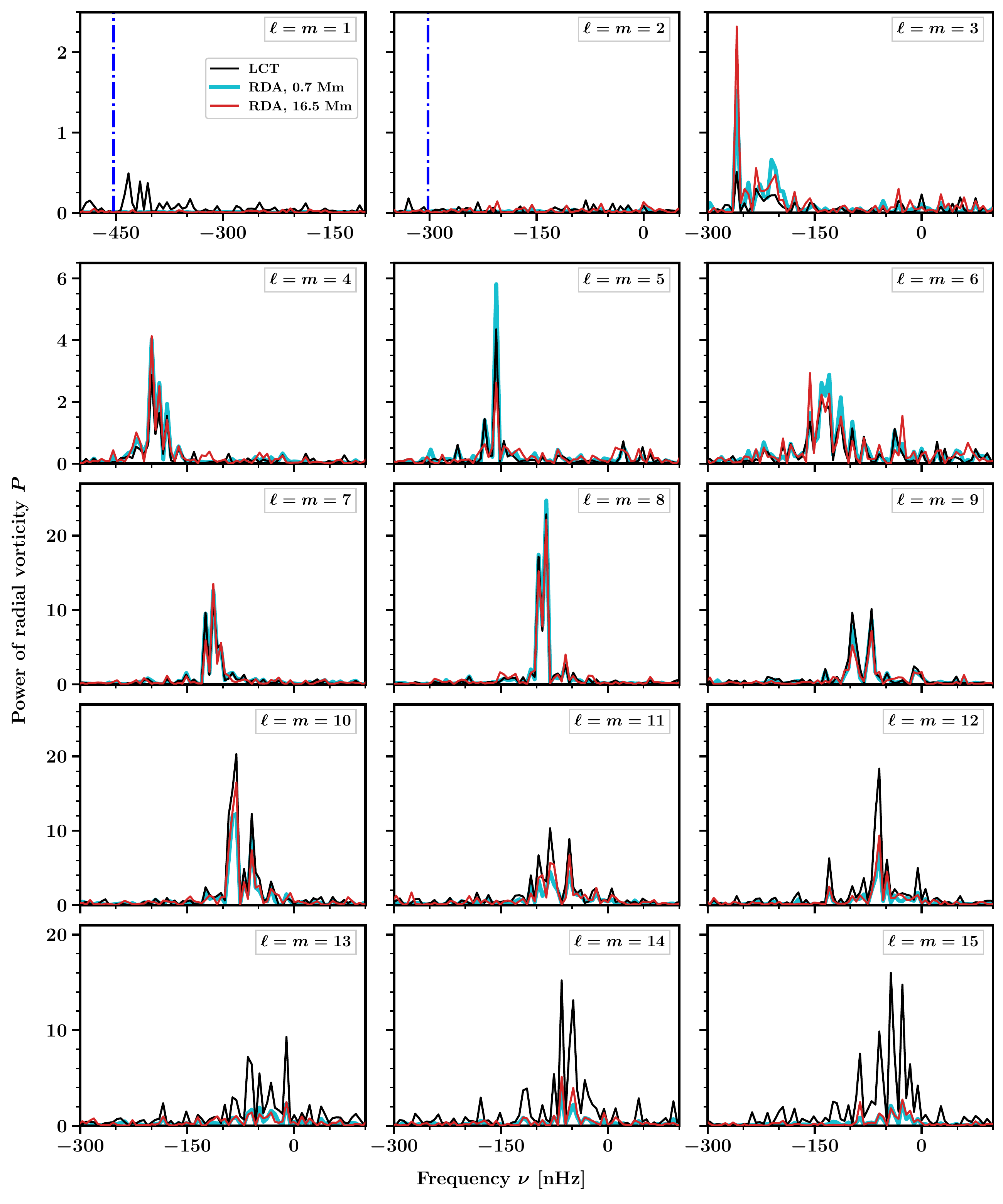}
\caption{
Sectoral power spectra ($\ell = m$) showing the Rossby wave power in the LCT data (black line) and the RDA data at depths $0.7$ and $\SI{16.5}{\mega\metre}$ (cyan and red lines). The power is normalized by the average of the $m = 8$ power in the range [$-300,100$]~nHz. The dash-dotted vertical lines in the $m = 1$ and $m = 2$ panels indicate the frequencies $\omega = -2\Omega_{\text{eq}}/(m + 1)$. The frequency axes of the $m = 1$ and $m = 2$ panels are shifted with respect to the other panels.
}
% \vspace{1cm}
\label{fig_power_spectrum_all_mvals}
\end{figure*}

Figure~\ref{fig_power_spectrum_all_mvals} shows the power versus frequency for different azimuthal orders $m$. We divide the power, at each $m$, by the frequency average of the $\ell = m = 8$ mode over [$-300,100$]~nHz near the surface ($\SI{0.7}{\mega\metre}$). The power decreases from $0.7$ to $\SI{9.9}{\mega\metre}$, then increases toward $\SI{16.5}{\mega\metre}$, but the depth dependence is modest ($\leq \SI{20}{\percent}$). We also see that the wave power decreases with $m$ faster for RDA than for LCT owing to the different sensitivity kernels, as found by \citetalias{Loeptien2018}. The $\ell = m = 3$ signal has a multi-peak structure and is thus difficult to measure. We do not observe Rossby waves for $\ell = m \leq 2$; the dash-dotted blue lines for $\ell = m = 1$ and $\ell = m = 2$ indicate the expected mode frequencies from the textbook dispersion relation.

The wave power side lobes due to the window function explain why the $\ell = m = 6$ side lobe in Fig.~\ref{fig_power_spectrum_2d} even exceeds the main signal: the adjacent $\ell = m = 7$ mode has a higher relative power (see Fig.~\ref{fig_power_spectrum_all_mvals}). Systematic effects that are fixed in the Stonyhurst reference frame can be easily misinterpreted as an $\ell = m = 1$ Rossby wave signal (see the LCT curve in Fig.~\ref{fig_power_spectrum_all_mvals}), as their frequency (the rotation rate) is equal to the $\ell = m = 1$ Rossby wave frequency.

We assume that there is background power contributing to the observed power at the Rossby peak, but measuring its contribution directly at the peak is impossible. Since we are limited by the side lobes, we use a region halfway between the peak and the next side lobe, i.e., shifted from the peak by half the rotation rate. We checked that the shift direction does not matter much, so for the central background frequency, we use the Rossby wave frequencies $\nu_0^{\text{ref}}$ from \citetalias{Liang2019} and \citetalias{Loeptien2018} for $m = 3$ and $m \geq 4$, respectively, plus half the rotation frequency $\nu_{\text{eq}}$. We use the full widths at half maximum $\gamma^{\text{ref}} = \Gamma^{\text{ref}}/2\pi$ from \citetalias{Liang2019} and \citetalias{Loeptien2018} for $m = 3$ and $m \geq 4$, respectively, and perform a least-squares second-order polynomial fit in $m$ to obtain a smoothed linewidth $\gamma_{\text{smooth}}$. We use $\gamma_{\text{smooth}}$ for the width of the peak and background frequency intervals. Thus our peak and background frequency intervals at each $m$ are
\begin{equation}
\label{eq_freqintvals}
\begin{aligned}
&\text{peak interval:} \quad &&\nu_0^{\text{ref}} &&\pm \gamma_{\text{smooth}}, \\
&\text{background interval:} \quad &&(\nu_0^{\text{ref}} + \nu_{\text{eq}}/2) &&\pm \gamma_{\text{smooth}}.
\end{aligned}
\end{equation}
These definitions are used in the analysis of latitudinal eigenfunctions in Sect.~\ref{sect_latfunc}.

In Fig.~\ref{fig_power_spectrum_2d}, we see that the peak interval (dashed red lines) typically captures the main wave power well. The 1D power spectra, however, reveal that the background interval (not shown in Fig.~\ref{fig_power_spectrum_all_mvals}, but see dashed black lines in Fig.~\ref{fig_power_spectrum_2d}), however, is potentially contaminated by scattered signal power, for example, for $\ell = m = 5$, $6$, and $14$. To check how the frequency interval definition affects our results, we performed our analysis for several different peak and background intervals. The results are consistent, thus we adopt Eq.~\ref{eq_freqintvals} for the peak and background intervals.

Unlike \citetalias{Loeptien2018}, we see evidence for non-sectoral components of the Rossby waves. For $\ell = m + 2$ the 2D power spectrum shows, for $6 \leq m \leq 13$, a ridge of power at very similar frequencies to those of the $\ell = m$ Rossby waves seen in Fig.~\ref{fig_power_spectrum_2d}, apart from a higher relative noise level and side lobes. This is confirmed by the 1D cuts at fixed values of $m$. We do not see structure in the power spectra for $\ell = m + k$ other than for $k = 0$ and $k = 2$. In Sect.~\ref{sect_latfunc_results}, we indeed show that the latitudinal eigenfunctions of Rossby waves are not sectoral spherical harmonic functions (in agreement with \citetalias{Loeptien2018}).

\subsection{Latitudinal eigenfunctions of Rossby waves}
\label{sect_latfunc}

To estimate the latitudinal eigenfunctions, we first remove small-scale convection from the LCT maps via smoothing with a $\SI{6}{\degree}$ Gaussian in latitude. Next we compute the Fourier transform of the radial vorticity maps $\zeta(t,r,\lambda,\varphi)$ from LCT and RDA in time and longitude as follows:
\begin{equation}
\hat{\zeta}_m(\nu,r,\lambda) = \sum_{t} \sum_{\varphi} \zeta(t,r,\lambda,\varphi) e^{i (2\pi \nu t - m\varphi)}.
\end{equation}
The variables are discrete and take values at time steps $t_j = j T/N_t$ (integer $0 \leq j < N_t$), longitudes $\varphi_k = 2\pi k/N_\varphi$ (integer $0 \leq k < N_\varphi$), frequencies $\nu_s = s/T$ (integer $s$, with $-N_t/2 \leq s \leq N_t/2 - 1$ for even $N_t$), and azimuthal orders $m$ (integer, with $-N_\varphi/2 \leq m \leq N_\varphi/2 - 1$ for even $N_\varphi$). In this case, $T$, $N_t$, and $N_\varphi$ are the observation period and the number of data points in time and longitude, respectively. We apply a filter to select the Rossby waves one $m$ at a time, i.e.,
\begin{equation}
\bar{\zeta}_m(\nu,r,\lambda) = \hat{\zeta}_m(\nu,r,\lambda) F_m(\nu).
\end{equation}
The filter $F_m(\nu)$ is equal to one within the Rossby wave ridge and zero elsewhere. Since $\zeta(t,r,\lambda,\varphi)$ is real, the symmetry $\bar{\zeta}_m(\nu,r,\lambda) = \bar{\zeta}_{-m}^*(-\nu,r,\lambda)$ applies. We then transform back to time to obtain
\begin{equation}
\tilde{\zeta}_m(t,r,\lambda) = \frac{1}{N_t} \sum_{\nu} \bar{\zeta}_m(\nu,r,\lambda) e^{-i 2\pi \nu t}.
\end{equation}
In this way we obtain filtered time-latitude vorticity maps for every $m$. Because there is no symmetry $\bar{\zeta}_m(\nu,r,\lambda) = \bar{\zeta}_m^*(-\nu,r,\lambda)$, the filtered vorticity maps $\tilde{\zeta}_m(t,r,\lambda)$ are in general complex.

\citetalias{Loeptien2018} do a similar analysis for LCT data, in particular for rotation-averaged maps and filtering within $\pm\SI{30}{\nano\hertz}$ around the central mode frequencies. We do the entire latitudinal eigenfunction analysis for LCT and RDA, for full time-resolution maps and maps averaged in time within individual solar rotations, and for a $\pm\SI{27}{\nano\hertz}$ (five frequency pixels) and a linewidth filter (Eq.~\ref{eq_freqintvals}) around the central mode frequencies. The different time-resolution and filtering cases yield consistent results; we thus show only the outcome for the full time-resolution and linewidth filtering. However, \citetalias{Loeptien2018} take the real part of the complex $\tilde{\zeta}_m(t,r,\lambda)$. This is equivalent to assuming that the phase of the eigenfunction is independent of latitude. We address the implications of this in Sect.~\ref{sect_latfunc_results} in more detail.

To estimate uncertainties for all results in this paper, we split the data into equal-size time intervals, apply our analysis to each chunk, and calculate the standard deviation over the results (for complex quantities separately for the real and imaginary part). Appendix~\ref{app_error_estimation_validation} gives more details on error estimation and validation. Because of the small number of chunks, low-number statistics are an issue and the reported error bars are relatively uncertain. 

For the sake of clarity, for the simple case of a single $m$ Rossby wave with a single frequency $\nu_m$ and an eigenfunction $C_m(r,\lambda)$, the vorticity field would be given by
\begin{equation}
\zeta_m(t,r,\lambda,\varphi) \propto \textrm{Re}\left( C_m(r,\lambda) e^{i (m \varphi - 2\pi \nu_m t)} \right).
\end{equation}
We apply two different methods to obtain the eigenfunctions $C_m(r,\lambda)$ near the surface, the covariance method (Sect.~\ref{sect_latfunc_cov}), and the SVD method (Sect.~\ref{sect_latfunc_svd}). The former is used also by \citetalias{Loeptien2018}.

\subsubsection{Covariance}
\label{sect_latfunc_cov}

We calculate, at each $m$, the temporal covariance of the vorticity $\tilde{\zeta}$ between the equator near the surface (target depth $R = R_\odot - \SI{0.7}{\mega\metre}$ for RDA) and all other latitudes and depths, normalized by the variance at the equator near the surface
\begin{equation}
\label{eq_eigenfunc_cov_covnorm}
C_m(r,\lambda) = \frac{\langle \tilde{\zeta}^{'}_{m}(t,r,\lambda) \tilde{\zeta}_m^{'*}(t,r = R,\lambda = \SI{0}{\degree}) \rangle_t}{\langle \vert \tilde{\zeta}_m^{'}(t,r = R,\lambda = \SI{0}{\degree}) \vert^2 \rangle_t}, 
\end{equation}
where the angle brackets $\langle \cdot \rangle_t$ denote a temporal average and $\tilde{\zeta}^{'} = \tilde{\zeta} - \langle \tilde{\zeta} \rangle_t$ is the centered vorticity. The function $C_m(r,\lambda)$ is complex-valued since $\tilde{\zeta}_m$ is in general complex. By construction $C_m(r = R,\lambda = \SI{0}{\degree})$ is unity. Appendix~\ref{app_latitudinal_eigenfunction_methods} shows that $C_m$ can also be obtained by a linear fit to the vorticity. The same covariance can be computed with the LCT data.

\subsubsection{Singular value decomposition}
\label{sect_latfunc_svd}

We present a second, new method to obtain latitudinal eigenfunctions. We want to separate the filtered vorticity at each azimuthal order $m$ and depth $r$, i.e., a 2D matrix, into a latitude and a time dependence, i.e.,
\begin{equation}
\label{eq_svd_concept}
\tilde{\zeta}_m(t,r,\lambda) \propto f_m(t) C_m(r,\lambda).
\end{equation}
Applying a singular value decomposition (SVD), we can decompose the vorticity as
\begin{equation}
\tilde{\zeta}_m(t,r,\lambda) = \sum_{j = 0}^{k-1} s_{(r,\,m),\,j} U_{(r,\,m),\,j}(t) V_{(r,\,m),\,j}(\lambda),
\end{equation}
where $s_{(r,\,m),\,j}$ is the singular value of index $j$ with left and right singular vectors $U_{(r,\,m),\,j}$ and $V_{(r,\,m),\,j}$ and $k$ is the minimum between the number of grid points in time and latitude. The square of $s_{(r,\,m),\,j}$ measures the variance captured by its singular vectors. By convention the singular values are sorted in descending order, thus the first singular vector contains more variance than any other individual singular vector.

Assuming that there is only one nonzero singular value, $s_{(r,\,m),\,0}$, the SVD gives the desired decomposition of the vorticity into one time and one latitude function. Our observations indeed have one clearly dominant singular value.

Given that the noise at high latitudes increases steeply, we crop our vorticity maps for the SVD to latitudes within $\pm\SI{50}{\degree}$ of the equator. Also, the SVD does not account for the varying noise of the remaining latitudes. To ensure that latitudes with larger uncertainties are given less weight, we filter the original vorticity maps once more in Fourier space for the noise, calculate the temporal standard deviation $\sigma_m$ of the noise-filtered maps, and compute $\tilde{\zeta}_{\text{nw},\,m}(t,r,\lambda) = \tilde{\zeta}_m(t,r,\lambda)/\sigma_m(r,\lambda)$. We filter for the noise by taking either all frequencies except for five pixels around the peak or all frequencies within the background interval (see Eq.~\ref{eq_freqintvals}). The two different filters give consistent results. At each $m$, the SVD is performed on the weighted maps $\tilde{\zeta}_{\text{nw},\,m}$ and the resulting latitude vectors are multiplied by $\sigma_m$ again to undo the weighting. We apply the weighting only to LCT, since the ring-diagram data are already apodized (see Sect.~\ref{sect_data_methods}). We select the first latitude singular vector near the surface and normalize it by its value at the equator.

\subsubsection{Results for the latitudinal eigenfunctions}
\label{sect_latfunc_results}

\begin{figure*}
\centering
\includegraphics[width=\hsize]{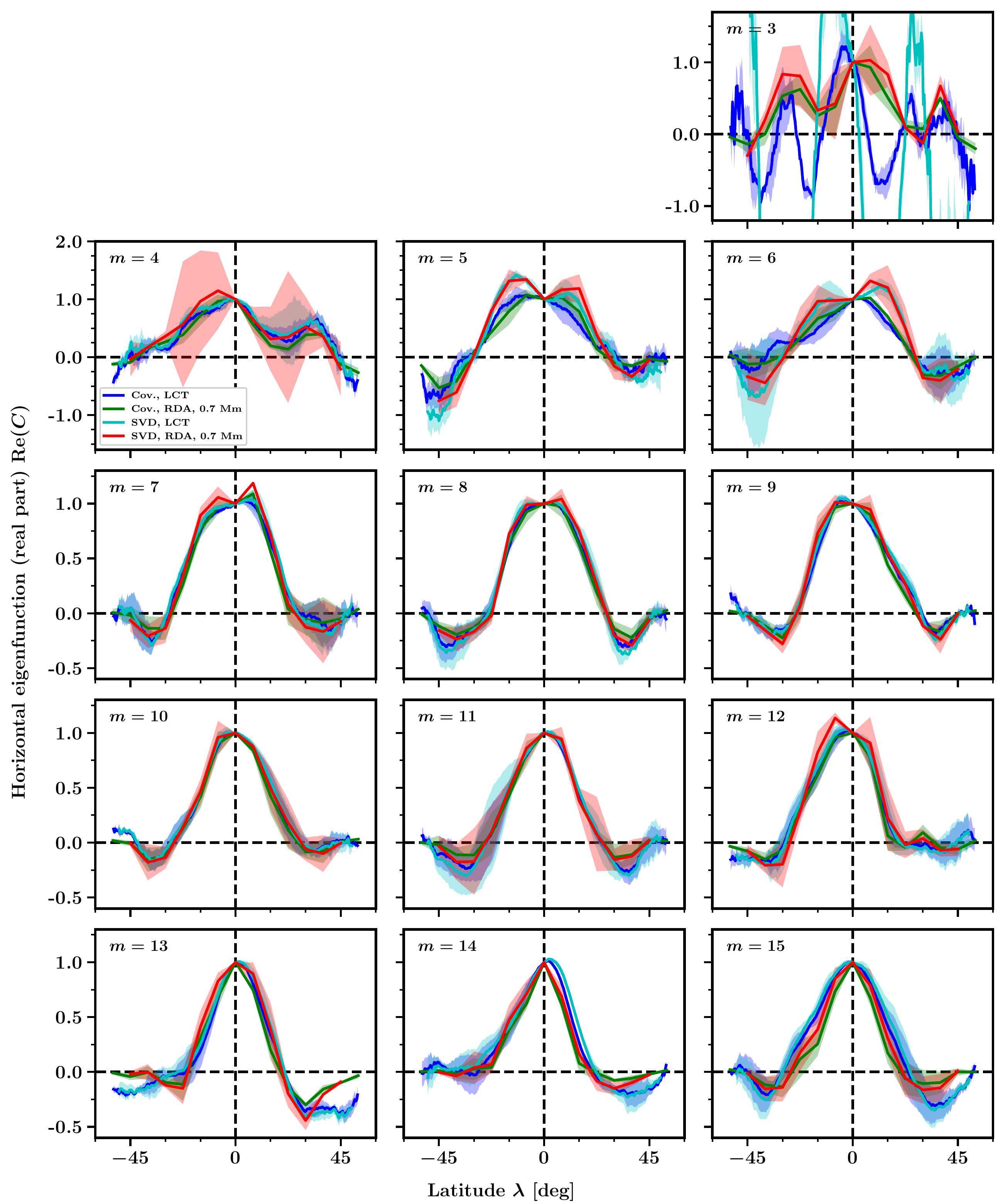}
\caption{
Real part of $C_m(\lambda)$ for different azimuthal orders $m$ and four different methods (see legend in panel $m = 4$). The shaded areas indicate the 1$\sigma$ error estimates. 
}
\vspace{1cm}
\label{fig_horiz_eigenfunction_all_mvals_real}
\end{figure*}

\begin{figure*}
\centering
\includegraphics[width=\hsize]{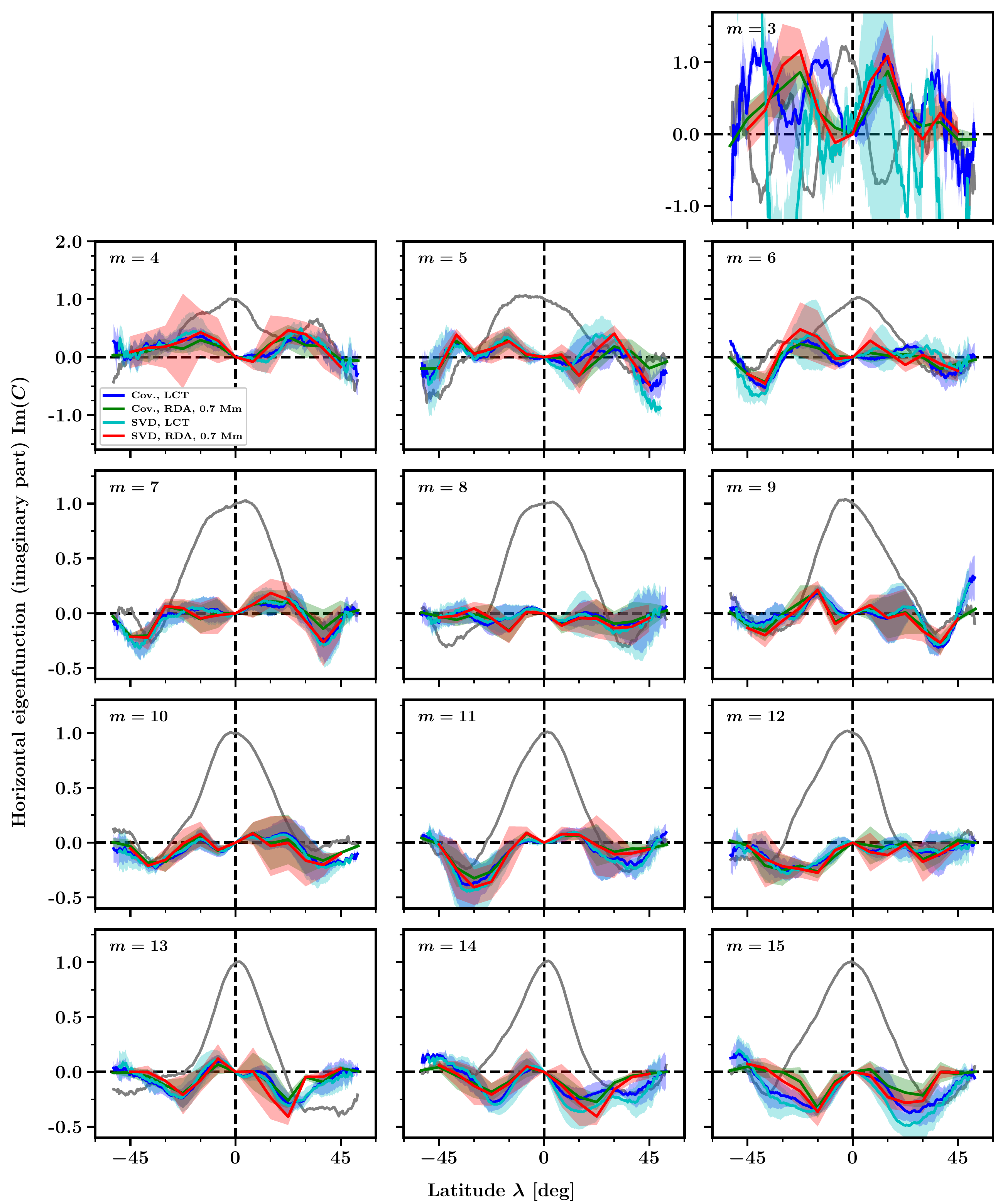}
\caption{
Imaginary part of $C_m(\lambda)$ for different azimuthal orders $m$ and four different methods (see legend in panel $m = 4$). The shaded areas indicate the 1$\sigma$ error estimates. For comparison, the solid gray curves show the real part of $C_m$ for the LCT covariance-based data. The plotting ranges are the same as in Fig.~\ref{fig_horiz_eigenfunction_all_mvals_real}.
}
\vspace{1cm}
\label{fig_horiz_eigenfunction_all_mvals_imag}
\end{figure*}

Figures~\ref{fig_horiz_eigenfunction_all_mvals_real} and \ref{fig_horiz_eigenfunction_all_mvals_imag} show the real and imaginary parts of the horizontal eigenfunctions of Rossby waves versus latitude for different $m$. The real part is consistent with the findings from \citetalias{Loeptien2018}. The imaginary part, however, was not discussed by \citetalias{Loeptien2018}.

In the current paper, we find that the LCT and the RDA results are mostly consistent for the near-surface layers. Also, almost all $m$ show agreement between the covariance and SVD results. This in particular holds for the modes with the largest amplitudes, i.e., for $7 \leq m \leq 10$. On the other hand, the modes $m = 4$ and to a lesser extent $m = 15$, where Rossby wave measurements become difficult, display larger errors but nonetheless consistent results. The $m = 3$ results for the different techniques disagree and are noisy. The $m = 5$ and $m = 6$ results for the real part differ slightly between the covariance and SVD methods. While the covariance yields a real part of the eigenfunction quite similar to those of other modes, the SVD-based results show maxima around latitudes of $\pm 10$-$\SI{15}{\degree}$. Apparently, there the SVD picks up some variance that is uncorrelated with the equator. It is unclear whether it is just noise, or a real signal of a different kind of latitudinal eigenfunctions.

The eigenfunction shape is similar for different modes. The real part decreases away from the equator, flips sign, and approaches zero after going through a local minimum. The imaginary part is much noisier than the real part, as indicated by the error estimates. For most $m$, it is close to zero and flat near the equator, but reaches minima at high latitudes. The latitude of the minima appears to move toward the equator with increasing $m$.

As can be seen from, for example, the red curves in Fig.~\ref{fig_horiz_eigenfunction_all_mvals_imag}, the imaginary part appears to be mostly positive for $3 \leq m \leq 6$. For $7 \leq m \leq 9$ the sign of the imaginary part is unclear. For $10 \leq m \leq 15$, the imaginary part is predominantly negative. The presence of an imaginary part induces a phase for the latitudinal eigenfunctions that can be interpreted as a latitude-dependent shift of the sinusoid in longitude. A positive sign of the imaginary part means that the horizontal eigenfunctions at high latitudes are leading in the retrograde direction with respect to the equator. Conversely, a negative sign would indicate that the eigenfunctions at high latitudes are trailing with respect to the equator. This may provide important constraints on the theory of latitudinal eigenfunctions of Rossby waves.

\begin{figure}
\centering
\includegraphics[width=\hsize]{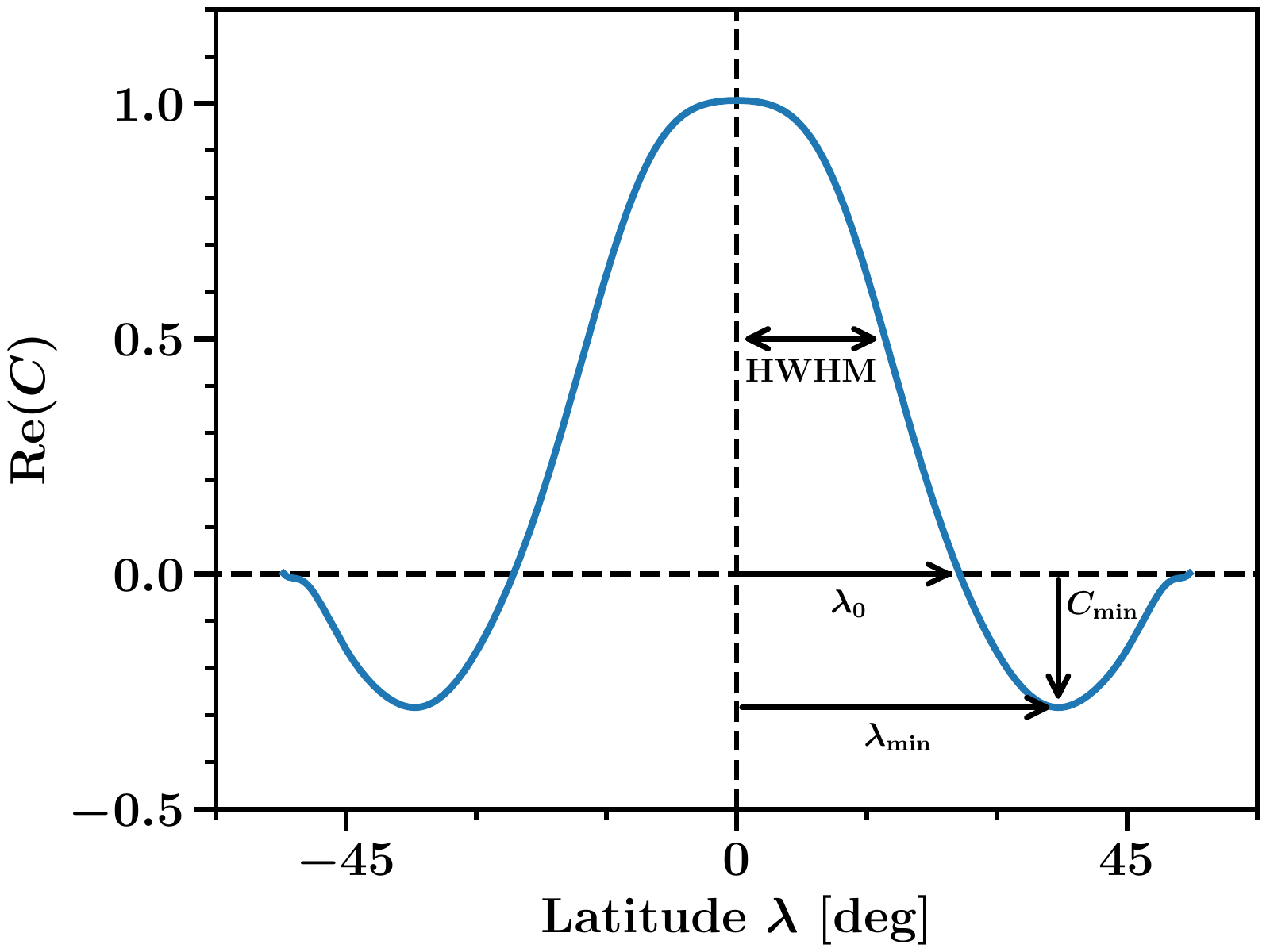}
\caption{
Schematic description of the real part of $C_m(\lambda)$ for a given $m$. The various parameters that describe the curve are the HWHM, the latitude at zero crossing ($\lambda_0$), the latitude at minimum ($\lambda_{\text{min}}$), and the minimum value ($C_{\text{min}}$).
}
\label{fig_horiz_eigenfunction_symm_m8}
\end{figure}

Figure~\ref{fig_horiz_eigenfunction_all_mvals_real} suggests that the real part of the eigenfunctions is more confined to low latitudes for higher values of $m$. We study the $m$-dependence of several characteristic parameters illustrated in Fig.~\ref{fig_horiz_eigenfunction_symm_m8}, i.e., the width at an eigenfunction real part of $\text{Re}(C) = 0.5$ (a half width at half maximum; HWHM), the latitude of the eigenfunction real part sign reversal, $\lambda_0$, and the latitude and value of the minimum, $\lambda_{\text{min}}$ and $C_{\text{min}}$, respectively. To reduce the noise level we derive the eigenfunctions from maps symmetrized in latitude before measuring these parameters.

To obtain the latitude widths at $\text{Re}(C) = 0.5$ and $\text{Re}(C) = 0$, we linearly fit the two points closest to these $\text{Re}(C)$ values. The latitude and value of the minimum are obtained by quadratically fitting three points around the minimum derived without fitting. We do this to avoid oscillating RDA results due to the coarse $\SI{7.5}{\degree}$ latitude sampling. For LCT, the effects of fitting the minimum (or not) are minimal. There are no results for $m = 3$ and $m = 4$ because of the poor quality and different shape of the eigenfunctions. We already stated the difficulties in characterizing these modes. As described at the beginning of Sect.~\ref{sect_latfunc} and in App.~\ref{app_error_estimation_validation}, to derive uncertainties, we compute the standard deviation over the results for different time chunks, separately for the real and the imaginary part.

\begin{table}
\caption{Parameters of the real part of $C_m(\lambda)$ for the LCT covariance-based data; see Fig.~\ref{fig_horiz_eigenfunction_symm_m8}. The parameters for $m = 3$ and $m = 4$ are not given owing to the large uncertainties.}
\label{table_params_eigenfunction}
\centering
\begin{tabular}{c c c c c}
\hline
\hline
$m$ & HWHM & $\lambda_0$ & $\lambda_{\text{min}}$ & $C_{\text{min}}$ \\
& [deg] & [deg] & [deg] & \\
\hline
5 & $20.7 \pm 2.8$ & $31.5 \pm 4.0$ & $46.5 \pm 3.6$ & $-0.38 \pm 0.17$ \\
6 & $16.1 \pm 2.8$ & $31.5 \pm 2.2$ & $44.0 \pm 3.3$ & $-0.24 \pm 0.13$ \\
7 & $18.7 \pm 0.4$ & $28.7 \pm 1.2$ & $37.0 \pm 4.2$ & $-0.17 \pm 0.07$ \\
8 & $17.1 \pm 0.8$ & $25.8 \pm 0.2$ & $36.5 \pm 1.8$ & $-0.28 \pm 0.06$ \\
9 & $16.0 \pm 1.3$ & $27.4 \pm 1.4$ & $35.2 \pm 0.4$ & $-0.17 \pm 0.02$ \\
10 & $14.7 \pm 1.1$ & $27.8 \pm 1.6$ & $34.5 \pm 2.1$ & $-0.11 \pm 0.06$ \\
11 & $14.3 \pm 1.2$ & $24.8 \pm 2.3$ & $34.9 \pm 2.4$ & $-0.25 \pm 0.09$ \\
12 & $13.8 \pm 1.4$ & $28.9 \pm 1.0$ & $39.6 \pm 1.8$ & $-0.14 \pm 0.01$ \\
13 & $11.3 \pm 2.0$ & $21.1 \pm 1.3$ & $47.1 \pm 2.2$ & $-0.31 \pm 0.02$ \\
14 & $12.0 \pm 0.8$ & $24.6 \pm 9.0$ & $35.5 \pm 5.5$ & $-0.11 \pm 0.07$ \\
15 & $14.7 \pm 2.0$ & $27.0 \pm 3.1$ & $37.7 \pm 2.0$ & $-0.26 \pm 0.04$ \\
\hline
\end{tabular}
\end{table}

Table~\ref{table_params_eigenfunction} shows how these parameters, measured for the LCT data from the covariance method, depend on $m$. Although not given in the table, we also measure the parameters for the RDA and the SVD results. We thus also discuss the $m$-dependence for those measurements; this dependence is mostly consistent with that of the LCT covariance-based parameters.

The latitude width at $\text{Re}(C) = 0.5$ indeed decreases with $m$, quasi-linearly between $m = 7$ and $m = 13$. The slope is roughly $\SI{-1}{\degree}$ per $m$. The decrease might flatten off at high $m$, but this could also be caused by noise. We observe slightly different latitude widths between the covariance and SVD eigenfunctions at low $m$ for $\text{Re}(C) = 0.5$, but similar widths at $\text{Re}(C) = 0$. Toward higher $m$, $\lambda_0$ is consistent with a flat profile, until around $m = 13$ the eigenfunction widths become smaller. The latitude of the minimum, $\lambda_{\text{min}}$, shows an $m$-(in)dependence similar to $\lambda_0$. There is a strong discrepancy for $m = 13$ between LCT and RDA, indicating that this mode is not trivial to characterize. This could be caused by noise. To some extent we could already see this in the power spectrum in Fig.~\ref{fig_power_spectrum_all_mvals}, where the $m = 13$ linewidth is large compared to all other $m$. The error on $\lambda_{\text{min}}$ might be underestimated here, since as seen in the asymmetric eigenfunctions in Fig.~\ref{fig_horiz_eigenfunction_all_mvals_real} the minimum of the LCT data is more poorly defined for $m = 13$ than for other modes. Finally, the value of the minimum, $C_{\text{min}}$, is different between the different analysis methods at $m = 5$ and $m = 6$, as seen before. Otherwise, it is quasi $m$-independent and has at most a very mild increase with $m$ from $m = 7$ onward, which is likely covered by noise, however.

As mentioned before, the latitudinal eigenfunctions appear to have two nodes (zero crossings) at latitudes $\pm\lambda_0$. This in combination with Fig.~\ref{fig_power_spectrum_2d} and the subsequent discussion indicates that the eigenfunctions have significant contributions from $\ell = m$ and $\ell = m + 2$ components. To quantify these contributions, we project the symmetric eigenfunctions $C_m(\lambda)$ onto associated Legendre polynomials $P_{\ell}^{m}(\sin\lambda)$, to obtain the coefficients
\begin{equation}
c_{\ell m} = \frac{\pi}{2 N_\lambda} \sum_{\lambda} C_m(r = R,\lambda) P_{\ell}^{m}(\sin\lambda) \cos\lambda.
\end{equation}
The sum goes over all latitudes $\lambda = k \pi/N_\lambda$ (integer $-N_\lambda/2 \leq k < N_\lambda/2$), where $N_\lambda$ is the number of data points in latitude. The $P_{\ell}^{m}(\sin\lambda)$ are normalized such that $\pi/N_\lambda \sum_\lambda (P_{\ell}^{m}(\sin\lambda))^2 \cos\lambda = 2$. The associated Legendre polynomials used in the decomposition are not orthogonal over the limited observed latitude range. However, we do not expect this to be a problem since we see later in this section that the near-sectoral associated Legendre polynomials, whose amplitude is concentrated toward the equator, are the dominant contributions to the latitudinal eigenfunctions. Because of the symmetry of the eigenfunctions, only $c_{\ell m}$ with even $\ell - m \geq 0$ are nonzero. We find that the real and the imaginary parts of the eigenfunctions for almost all $m$ can be approximated well (within 1$\sigma$) when using only the contributions from $c_{\ell m}$ for $m \leq \ell \leq m + 6$, except for $m = 3$, which is very noisy. The approximation also does not work well at the high latitudes (beyond $\pm\SI{40}{\degree}$) for the real part (for some modes) and at the near-equatorial latitudes for the imaginary part (for almost all modes).

\begin{table*}
\caption{Coefficients $c_{\ell m}$ for the LCT covariance-based data. Each bracketed pair of numbers refers to the real and imaginary parts of $c_{\ell m}$. The numbers in italics are not significantly different from zero (zero within 1$\sigma$).}
\label{table_params_eigenfunc_legendre_coeffs}
\tiny
\centering
\begin{tabular}{c c c c c}
\hline
\hline
$m$ & $c_{m m}$ & $c_{m+2,\,m}$ & $c_{m+4,\,m}$ & $c_{m+6,\,m}$ \\
\hline
3 & $(+0.026,+0.417) \pm (0.018,0.105)$ & $(-0.155,\mathit{+0.089}) \pm (0.032,0.103)$ & $(+0.118,-0.108) \pm (0.075,0.074)$ & $(-0.117,\mathit{+0.004}) \pm (0.015,0.029)$ \\
4 & $(+0.457,+0.136) \pm (0.016,0.060)$ & $(-0.097,+0.061) \pm (0.027,0.017)$ & $(\mathit{-0.010},-0.066) \pm (0.036,0.030)$ & $(-0.071,\mathit{+0.019}) \pm (0.014,0.026)$ \\
5 & $(+0.478,\mathit{+0.037}) \pm (0.031,0.070)$ & $(-0.263,\mathit{-0.021}) \pm (0.086,0.067)$ & $(-0.059,\mathit{-0.106}) \pm (0.052,0.112)$ & $(\mathit{+0.041},\mathit{-0.011}) \pm (0.045,0.040)$ \\
6 & $(+0.411,+0.021) \pm (0.055,0.019)$ & $(-0.202,\mathit{-0.050}) \pm (0.025,0.057)$ & $(\mathit{-0.005},-0.118) \pm (0.040,0.032)$ & $(\mathit{-0.006},+0.059) \pm (0.007,0.018)$ \\
7 & $(+0.473,\mathit{+0.026}) \pm (0.024,0.028)$ & $(-0.183,-0.055) \pm (0.015,0.024)$ & $(\mathit{-0.012},-0.073) \pm (0.012,0.023)$ & $(+0.043,+0.018) \pm (0.006,0.012)$ \\
8 & $(+0.441,\mathit{-0.013}) \pm (0.015,0.022)$ & $(-0.231,-0.038) \pm (0.021,0.027)$ & $(-0.047,\mathit{-0.009}) \pm (0.031,0.068)$ & $(+0.038,\mathit{+0.001}) \pm (0.017,0.027)$ \\
9 & $(+0.434,\mathit{+0.028}) \pm (0.013,0.042)$ & $(-0.162,-0.032) \pm (0.023,0.014)$ & $(\mathit{-0.010},-0.053) \pm (0.018,0.044)$ & $(+0.025,\mathit{+0.022}) \pm (0.005,0.030)$ \\
10 & $(+0.423,\mathit{+0.008}) \pm (0.024,0.012)$ & $(-0.139,-0.057) \pm (0.017,0.041)$ & $(\mathit{+0.011},-0.076) \pm (0.034,0.028)$ & $(+0.020,-0.013) \pm (0.001,0.010)$ \\
11 & $(+0.400,-0.010) \pm (0.027,0.003)$ & $(-0.173,-0.117) \pm (0.049,0.043)$ & $(-0.043,-0.064) \pm (0.027,0.058)$ & $(\mathit{+0.002},+0.024) \pm (0.018,0.024)$ \\
12 & $(+0.419,-0.040) \pm (0.029,0.023)$ & $(-0.117,-0.078) \pm (0.006,0.011)$ & $(-0.016,\mathit{-0.021}) \pm (0.010,0.032)$ & $(\mathit{-0.027},\mathit{-0.007}) \pm (0.042,0.030)$ \\
13 & $(+0.345,\mathit{-0.016}) \pm (0.036,0.023)$ & $(-0.195,-0.112) \pm (0.018,0.020)$ & $(-0.048,-0.028) \pm (0.029,0.011)$ & $(-0.071,+0.026) \pm (0.013,0.023)$ \\
14 & $(+0.380,-0.057) \pm (0.026,0.025)$ & $(-0.128,-0.102) \pm (0.038,0.044)$ & $(\mathit{-0.007},\mathit{+0.006}) \pm (0.027,0.022)$ & $(-0.019,\mathit{+0.001}) \pm (0.015,0.017)$ \\
15 & $(+0.431,-0.074) \pm (0.017,0.013)$ & $(-0.085,-0.151) \pm (0.054,0.029)$ & $(-0.072,\mathit{-0.041}) \pm (0.020,0.057)$ & $(-0.053,\mathit{-0.012}) \pm (0.036,0.020)$ \\
\hline
\end{tabular}
\end{table*}

Table~\ref{table_params_eigenfunc_legendre_coeffs} shows the coefficients $c_{\ell m}$ for $m \leq \ell \leq m + 6$ for the LCT covariance-based latitudinal eigenfunctions. As usual the uncertainties are calculated from the standard deviation over the coefficients for different time chunks (App.~\ref{app_error_estimation_validation}), separately for the real and the imaginary part. The real part of the eigenfunctions is clearly dominated by the $\ell = m$ component. The contribution from the $\ell = m + 2$ component is significant as well and has a relative strength of $30$-$\SI{50}{\percent}$. This is consistent with the observations from the 2D and 1D power spectra in Sect.~\ref{sect_power_spectra_rossby}. The real part of the $c_{m m}$ and $c_{m+2,\,m}$ each depend weakly on $m$. The real part of several of the coefficients with larger $\ell$ is insignificant. The imaginary part, on the other hand, has significant, dominant contributions at $\ell = m + 4$ for $m \leq 10$ and at $\ell = m + 2$ for $m \geq 11$, whereas the $\ell = m$ and $\ell = m + 6$ components are often insignificant. The term insignificant refers to an absolute value of $c_{\ell m}$ of less than 1$\sigma$. Nonetheless, independent of the estimated error bars, 11 out of 12 modes within $4 \leq m \leq 15$ have the same sign for $c_{m+4,\,m}$, suggesting that the $\ell = m + 4$ contribution to the real part is significant, despite being within $1\sigma$ from zero. A similar argument holds for the imaginary part of the latitude dependence of the Rossby wave eigenfunctions.

For the latitudinal eigenfunctions of Rossby waves there are so far only a few theoretical studies such as \citet{Lee1997} and \citet{Townsend2003}. These studies typically gave either analytic (asymptotic) expressions and/or numeric calculations, but the former expressions do not agree well with the latter calculations for Rossby waves \citep{Townsend2003}. Although both studies indicate that the latitudinal eigenfunctions are not concentrated near the equator, we cannot sensibly compare their findings to our measurements. In particular these models assume a uniform rotation rate. Also these authors used the traditional approximation, i.e., they neglected the horizontal component of the rotation vector. This approximation requires the squared Brunt-V{\"a}is{\"a}l{\"a} frequency $N^2$ to be much higher than both the squared oscillation frequency $\omega^2$ and the squared rotation rate $\Omega^2$. The validity of the traditional approximation thus has to be critically examined within the convection zone of the Sun.

\subsection{Radial eigenfunctions of Rossby waves}

\subsubsection{Depth-dependent ring-diagram systematics}

To study the Rossby wave depth dependence, we must check to which depths RDA is reliable. For this we compare the solar rotation profile from ring-diagram velocities with the results from SDO/HMI global modes from the JSOC data series hmi.V\_sht\_2drls \citep{Larson2018}. The global modes have a $72$-day time sampling from April 30, 2010 to June 4, 2017, a $\SI{1.875}{\degree}$ latitude sampling, and a nonlinear depth grid with many more points near the surface than at larger depths. Global modes are expected to give a precise and accurate solar rotation profile. We interpolate the global mode results to the ring-diagram latitude-depth grid via 2D bicubic splines, which is reasonable because the global-mode inversions do not vary on scales of their original grid; we then average the $72$-day chunks over time. The chunk scatter of the rotation rate is used to estimate the uncertainty. We divide the ring-diagram data into five intervals of length $480$ time steps ($20$ rotations), average the chunks over time and estimate the error from the scatter, convert the velocities into rotation rates, and add the sidereal Carrington rate to correct for the ring-diagram tracking.

\begin{figure}
\centering
\includegraphics[width=\hsize]{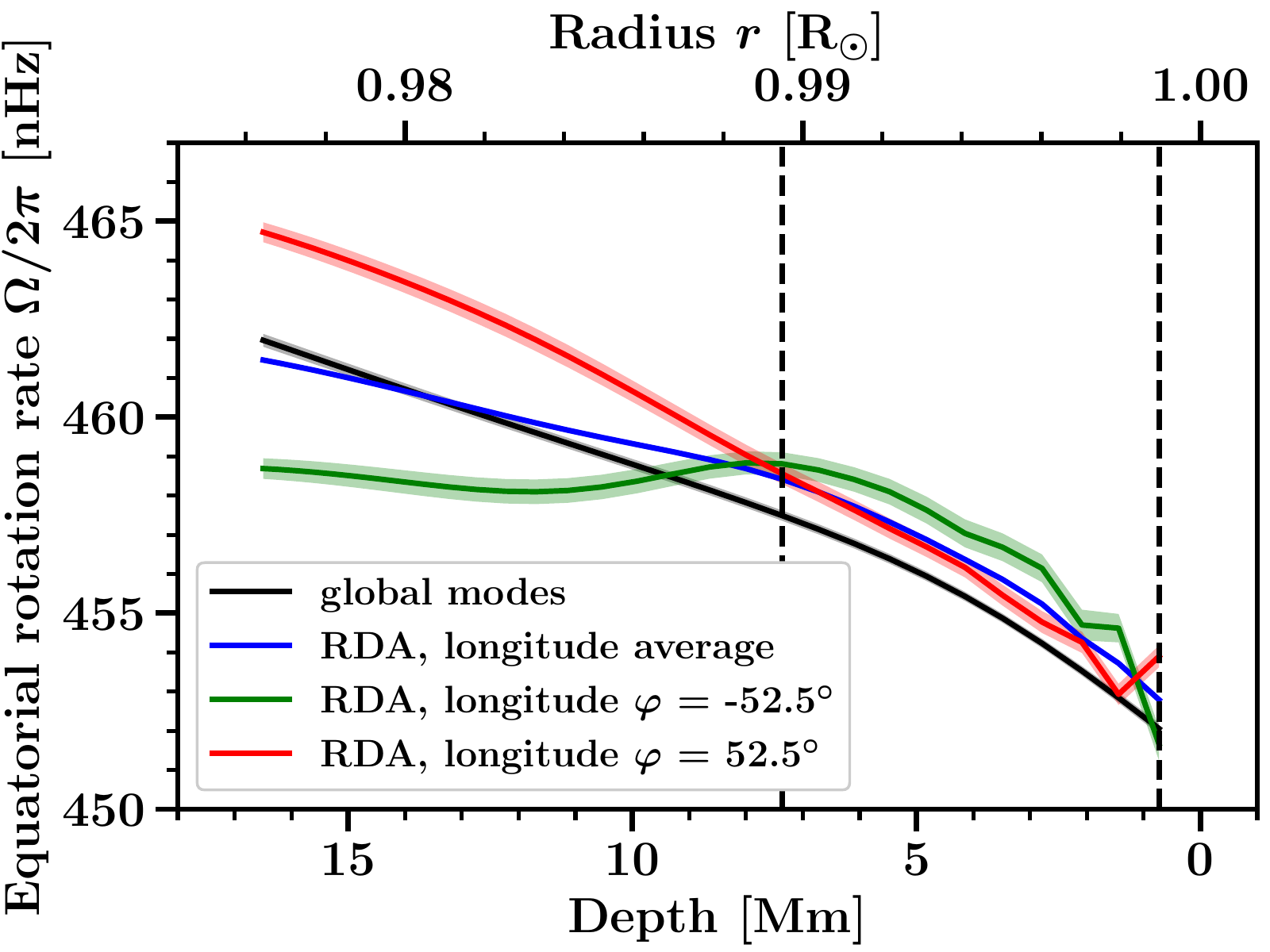}
\caption{
Solar equatorial rotation rate as a function of depth. The global-mode helioseismology result is given by the black curve. The blue curve is for RDA after averaging over all longitudes. The green and red curves show the ring-diagram results at Stonyhurst longitudes $\pm\SI{52.5}{\degree}$.The shaded areas give the 1$\sigma$ error estimates. The observations cover more than seven years. The dashed black lines indicate the depth range within which the ring-diagram results are in best agreement with each other. 
}
\label{fig_rotation_systematics}
\end{figure}

Figure~\ref{fig_rotation_systematics} shows the equatorial rotation rate versus depth from global modes and ring-diagram velocities, both averaged over longitude and at Stonyhurst longitudes of $\pm\SI{52.5}{\degree}$ (the outermost longitudes in our vorticity maps). The global modes yield a smooth profile with extremely small errors. The ring-diagram data show a small offset at small depths, but, more importantly, inconsistency with the global modes at large depths. Of course, it is difficult to judge how well the results should agree because of the different kernels of the datasets and thus different depth (and latitude) sensitivities. The $\SI{-52.5}{\degree}$ longitude curve has a small local maximum around $\SI{8}{\mega\metre}$. For longitudes even further east (not shown), the rotation rate has an even stronger excess (a bump) there.

The most worrisome point is the disagreement between different ring-diagram longitudes themselves and also with the longitude average, below roughly $\SI{8}{\mega\metre}$ (indicated by the left dashed line in Fig.~\ref{fig_rotation_systematics}). Because we averaged the data over more than seven years, any short-lived flows and even longer-lived structures should be filtered away and the longitude gradient from east to west should thus not exist. This points to a deeper problem with the ring fits and the pipeline processing that generated these fits. The presence of systematic effects in HMI ring-diagram data has also been extensively discussed in \citet{Komm2015}.

Finally, we note that Fig.~\ref{fig_rotation_systematics} is affected by an issue related to the ring-diagram inversion, since the inversion does not account for the quantity $\beta_{n \ell}$. A discussion of this issue and a brief check of the magnitude of the effect is given in App.~\ref{app_rda_inversion_issues}. The latter showed that the main effect is a depth-independent underestimation of the ring-diagram velocities by $1$-$\SI{2}{\metre\per\second}$ or equivalently of the rotation rate by less than $\SI{0.5}{\nano\hertz}$. This does not affect our main conclusions. The small, but significant difference between the rotation rates from global modes and ring diagrams cannot be caused by the $\beta_{n \ell}$ issue (it has the wrong sign), but may possibly instead be due to different averaging kernel widths, systematics, or other unknown effects.

\subsubsection{Determining the Rossby wave depth dependence}

In this section, we discuss only the sectoral ($\ell = m$) component of the power spectrum of radial vorticity. The Rossby wave power $P_m(\nu,r)$ and phase $\Phi_m(\nu,r)$ thus depend on frequency, depth, and azimuthal order. Based on the assumption of damped oscillations and stochastic wave excitation, we perform a maximum-likelihood Lorentzian fit \citep{Anderson1990} to the power spectra for the longer ring-diagram period, separately at each $m$. We use the functional form
\begin{equation}
\label{eq_power_fit_lorentz}
P_{\text{fit},\,m}(\nu,r) = \frac{A_m(r)}{4(\nu - \nu_{0,\,m})^2/\gamma_m^2 + 1} + B_m(r).
\end{equation}
We fit all the depths (except for the surface, i.e., $r = \SI{0.0}{\mega\metre}$, where the ring-diagram data are unreliable) at once, with a common central frequency $\nu_{0,\,m}$ and linewidth $\gamma_m$, but with individual amplitudes $A_m(r)$ and backgrounds $B_m(r)$. The Lorentzian fit of the power spectra, in most cases, fits well to the observations. As seen in Fig.~\ref{fig_power_spectrum_all_mvals}, the $\ell = m = 6$ and $\ell = m = 13$ modes have large linewidths and their power spectra show fine structure. The $\ell = m = 3$ mode has been fit by \citetalias{Liang2019}, but not by \citetalias{Loeptien2018}.

To determine error bars for the amplitudes and backgrounds via chunked data (App.~\ref{app_error_estimation_validation}), we also perform the Lorentzian fit for each chunk separately, fitting again all depths together, but keeping the central frequency and linewidth fixed at the fit results of $\nu_{0,\,m}$ and $\gamma_m$ from Eq.~\ref{eq_power_fit_lorentz} for the full time period (to prevent unstable fits). Because we keep these parameters fixed for each chunk, we cannot derive their uncertainties based on the standard deviation over the chunks. We thus do a Monte Carlo simulation and generate $1000$ realizations of synthetic power spectra according to Eq.~\ref{eq_monte_carlo_model} (App.~\ref{app_error_estimation_validation}) and perform the Lorentzian fit for each realization analogously to the fit for the observations. The median of the parameters over the Monte Carlo realizations is consistent with the fit parameters for the observations. While the error based on the Monte Carlo simulation contains realization noise, the model we use (Lorentzian and stochastic excitation) does not include all features of the observed power spectra. The chunk-based error likely describes the physical system more accurately, by also including other variance contributions, such as from temporal effects on the Rossby waves; we could imagine, for example, solar cycle effects. This may also explain the discrepancy between the two types of errors of order $\SI{30}{\percent}$ (App.~\ref{app_error_estimation_validation}). This disagreement is, however, small enough to not affect the significance of the results for the radial eigenfunctions. We thus use the uncertainties based on the Monte Carlo simulation for the central frequency and the linewidth.

\setlength{\extrarowheight}{2pt}
\begin{table}
\caption{Measured frequencies and linewidths of the Rossby waves from RDA sectoral power spectra with azimuthal orders in the range $3\leq m \leq 15$. Previous measurements (with superscript 'ref') are also listed for comparison.}
\label{table_params_lorentz_fit}
\centering
\begin{tabular}{c r@{}l r@{}l r@{}l r@{}l c}
\hline
\hline
 & \multicolumn{4}{c}{This work} & \multicolumn{5}{c}{Previous work} \\
\hline
$m$ & \multicolumn{2}{c}{$\nu_{0,\,m}$} & \multicolumn{2}{c}{$\gamma_m$} & \multicolumn{2}{c}{$\nu_{0,\,m}^{\text{ref}}$} & \multicolumn{2}{c}{$\gamma_{m}^{\text{ref}}$} & Ref. \\
& \multicolumn{2}{c}{[$\SI{}{\nano\hertz}$]} & \multicolumn{2}{c}{[$\SI{}{\nano\hertz}$]} & \multicolumn{2}{c}{[$\SI{}{\nano\hertz}$]} & \multicolumn{2}{c}{[$\SI{}{\nano\hertz}$]} & \\
\hline
$3$ & $-230$ & $^{+5}_{-4}$ & $40$ & $^{+13}_{-11}$ & $-253$ & $~\pm~2$ & $7$ & $^{+4}_{-3}$ & \citetalias{Liang2019} \\
$4$ & $-195$ & $~\pm~3$ & $16$ & $^{+7}_{-5}$ & $-194$ & $^{+5}_{-4}$ & $18$ & $^{+14}_{-7}$ & \citetalias{Loeptien2018}\\
$5$ & $-159$ & $^{+3}_{-2}$ & $12$ & $^{+6}_{-5}$ & $-157$ & $~\pm~4$ & $11$ & $^{+14}_{-6}$ & \citetalias{Loeptien2018}\\
$6$ & $-119$ & $~\pm~6$ & $84$ & $^{+22}_{-19}$ & $-129$ & $~\pm~8$ & $47$ & $^{+28}_{-16}$ & \citetalias{Loeptien2018} \\
$7$ & $-111$ & $~\pm~3$ & $20$ & $^{+7}_{-5}$ & $-112$ & $~\pm~4$ & $17$ & $^{+10}_{-7}$ & \citetalias{Loeptien2018} \\
$8$ & $-89$ & $~\pm~3$ & $19$ & $^{+7}_{-6}$ & $-90$ & $~\pm~3$ & $12$ & $^{+7}_{-5}$ & \citetalias{Loeptien2018} \\
$9$ & $-77$ & $~\pm~4$ & $40$ & $~\pm$ 11 & $-86$ & $~\pm~6$ & $37$ & $^{+21}_{-11}$ & \citetalias{Loeptien2018} \\
$10$ & $-77$ & $^{+4}_{-3}$ & $29$ & $^{+10}_{-7}$ & $-75$ & $~\pm~5$ & $28$ & $^{+12}_{-10}$ & \citetalias{Loeptien2018} \\
$11$ & $-64$ & $^{+4}_{-5}$ & $47$ & $^{+13}_{-12}$ & $-75$ & $~\pm~7$ & $43$ & $^{+23}_{-13}$ & \citetalias{Loeptien2018} \\
$12$ & $-59$ & $~\pm~4$ & $35$ & $^{+11}_{-9}$ & $-59$ & $~\pm~6$ & $42$ & $^{+20}_{-12}$ & \citetalias{Loeptien2018} \\
$13$ & $-45$ & $~\pm~6$ & $76$ & $^{+22}_{-20}$ & $-40$ & $~\pm~10$ & $71$ & $^{+38}_{-22}$ & \citetalias{Loeptien2018} \\
$14$ & $-47$ & $~\pm~5$ & $40$ & $^{+13}_{-11}$ & $-56$ & $^{+6}_{-7}$ & $36$ & $^{+20}_{-13}$ & \citetalias{Loeptien2018} \\
$15$ & $-39$ & $^{+5}_{-4}$ & $41$ & $^{+12}_{-11}$ & $-47$ & $^{+7}_{-6}$ & $40$ & $^{+21}_{-12}$ & \citetalias{Loeptien2018} \\
 \hline
\end{tabular}
\end{table}
\setlength{\extrarowheight}{0pt}

Table~\ref{table_params_lorentz_fit} compares the fit parameters from this study with the results from \citetalias{Liang2019} and \citetalias{Loeptien2018}. As in \citetalias{Liang2019} and \citetalias{Loeptien2018}, the upper and lower errors give the difference between the quantiles comprising the central $\SI{68.3}{\percent}$ (the distributions are non-Gaussian) and the fit parameters for the observations. We also calculate the uncertainties on the central frequency following \citet{Libbrecht1992} and find that those (symmetric) errors typically underestimate the Monte Carlo quantile errors by roughly $\SI{1}{\nano\hertz}$. A possible reason for this could be that we use a finite frequency fitting interval.

The fit parameters for the observations and those from \citetalias{Liang2019} and \citetalias{Loeptien2018} typically agree within 1$\sigma$ or better. The central frequencies and the linewidths for the $\ell = m = 6$ mode differ by $10$ and $\SI{37}{\nano\hertz}$, but the fit is sensitive to the fitting range. The $\ell = m = 3$ fit parameters do not agree. In \citetalias{Liang2019}, the authors, using $21$~years of data, observed that the multi-peak structure of the $\ell = m = 3$ power spectrum (Fig.~\ref{fig_power_spectrum_all_mvals}) seen in data with shorter periods collapses to a narrow single peak, which indicates that the discrepancy of the fit parameters is explained by stochastic excitation of the Rossby waves and vanishes when fitting data with a longer time period. Our errors are typically more symmetric and often smaller than those of \citetalias{Liang2019} and \citetalias{Loeptien2018}. The lower errors for the linewidth often agree better than the upper errors, indicating a tail of high values (skewness) in the \citetalias{Liang2019} and \citetalias{Loeptien2018} estimate distributions. Reasons for the differences in the error estimates may lie in the simultaneous fitting of all depths at once or in the different observation periods of our datasets and those of \citetalias{Loeptien2018}.

To determine the power depth dependence, we use the amplitude $A$ derived from the Lorentzian fit (see Eq.~\ref{eq_power_fit_lorentz}) and we define the normalized power of the signal as
\begin{equation}
\label{eq_power_normalized_lorentz}
\mathcal{P}_{\text{signal},\,m}(r) = \frac{A_m(r)}{\langle A_m(r) \rangle_r}.
\end{equation}
We thus normalize by the depth average of the amplitude of the Lorentzian. The normalized power is independent of temporal amplitude variations due to Rossby wave excitation.

\subsubsection{Results for the radial eigenfunctions}

\begin{figure}
\centering
\includegraphics[width=\hsize]{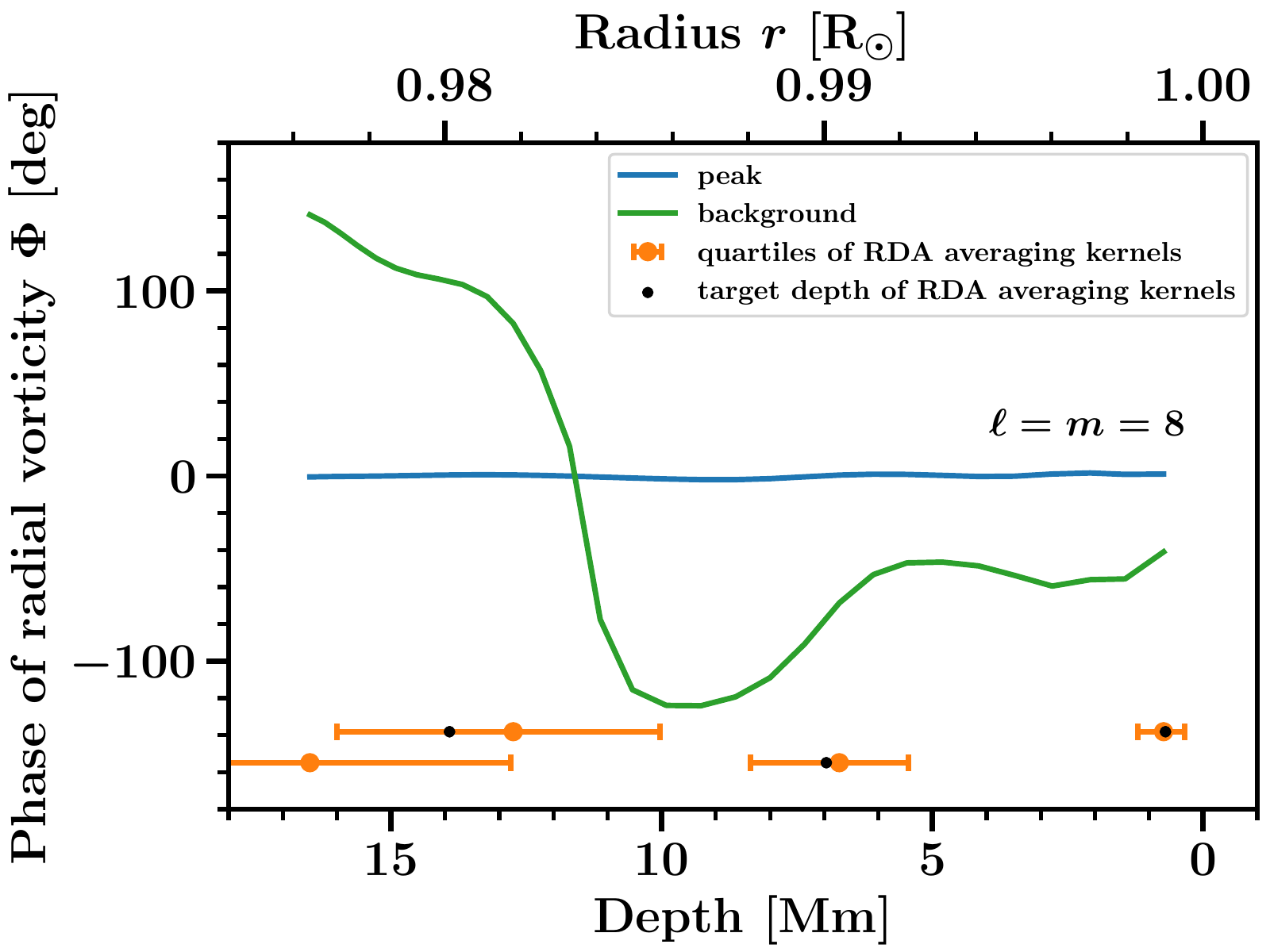}
\caption{
Phase at a frequency of roughly $\SI{-87.4}{\nano\hertz}$ corresponding to the peak of power for the Rossby mode $\ell = m = 8$, as a function of depth (blue curve). The green line shows the phase of the background at the center of the background interval (see Eq.~\ref{eq_freqintvals}). The depth refers to the median of the ring-diagram averaging kernels (orange dots), which corresponds to certain target depths (black dots).
}
\label{fig_phase_depth_m8}
\end{figure}

Figure~\ref{fig_phase_depth_m8} shows the depth dependence of the $\ell = m = 8$ phase, but the behavior is similar for other $m$. For easier comparison, we remove phase jumps of $\SI{360}{\degree}$ and move the depth average to zero. The phase at the frequency of maximum power is almost constant with depth, within roughly $\pm\SI{3}{\degree}$. The phase at the background, at the center of the background interval, varies much more strongly with depth, within roughly $\pm\SI{100}{\degree}$, although the phases at other background frequencies sometimes show much less variation. The background phase is not random in depth (see App.~\ref{app_error_estimation_validation}). In particular, phase changes are gradual and smooth; the depths are correlated. This could indicate a significant contribution from scattered signal power to the background. Nonetheless, peak and background display distinctly different depth dependences. We also find that phases at different frequencies across the peak and background are different. Frequency averages of phases are thus not useful. However, as seen, for single frequencies the phase at the peak is nearly constant with depth, while the background phase varies with depth.

Figure~\ref{fig_phase_depth_m8} also shows the main parameters of the ring-diagram averaging kernels for a few target depths, i.e., the first, second (median) and third quartiles and the width (interquartile range). The flow measurements are well-localized near the surface, but smeared out over a broad depth range at large depths. The ring-diagram depth covariance matrix (not shown) indicates a similar behavior and shows that different depths are mostly independent near the surface, while at the largest depths there is high correlation and they thus do not give independent results. This could maybe also explain why the background phase is not random in depth. At large depths, the center of the averaging kernels (second quartile) moves away from the target depth, but the averaging kernels are relatively symmetric.

The background power for different $m$ (not shown) generally increases with depth and at least for some modes there could be a minimum at $8$-$\SI{9}{\mega\metre}$, albeit with little significance given the large errors.

\begin{figure*}
\centering
\includegraphics[width=\hsize]{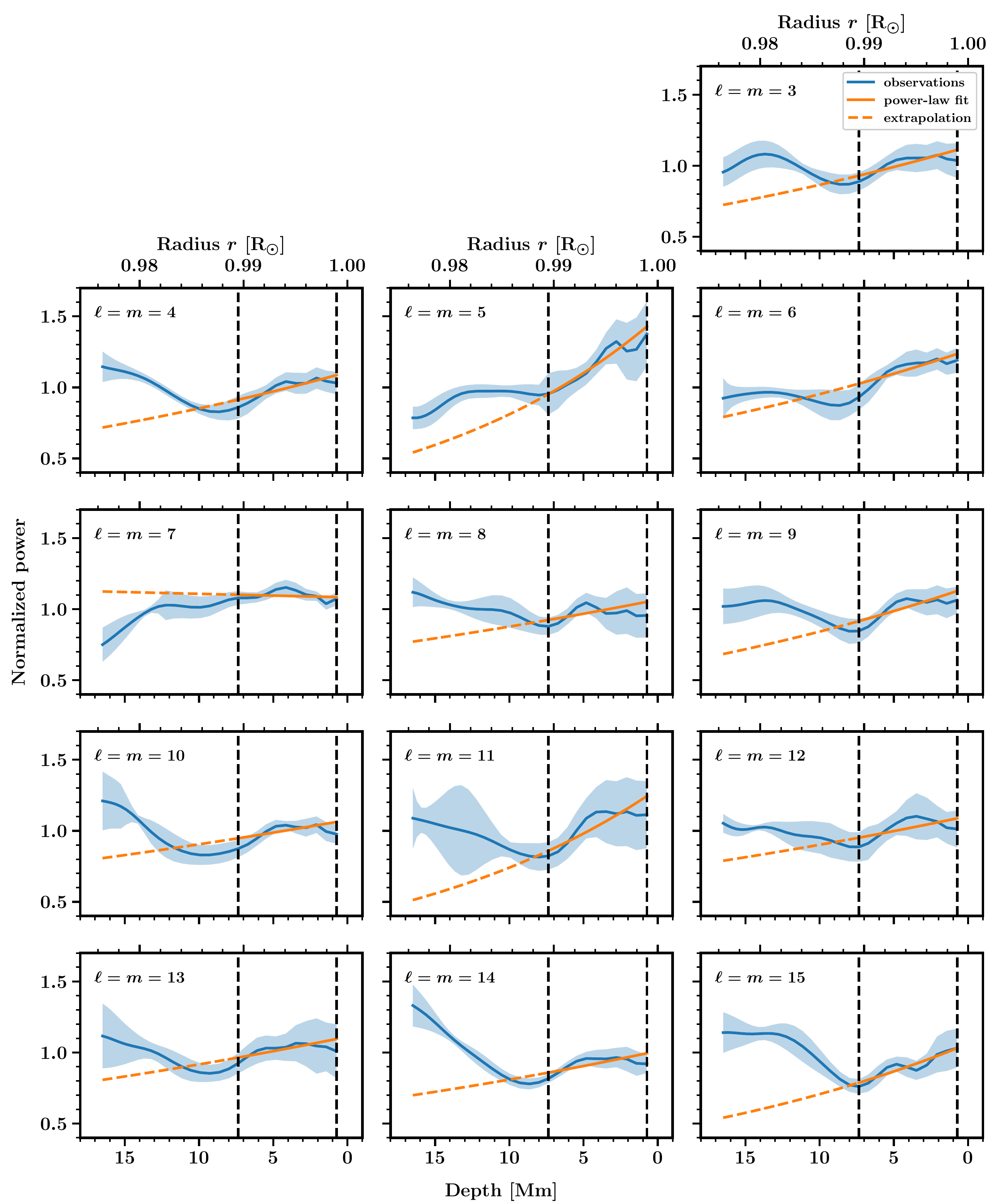}
\caption{
Blue lines show the Rossby wave power $\mathcal{P}_{\text{signal}}$ as a function of depth for different values of $m$. The blue shaded areas indicate the 1$\sigma$ errors. The orange curves are fits of the form $const. \times r^{2\alpha}$ over depths between $0.7$ and $\SI{7.4}{\mega\metre}$ (between the vertical dashed lines). The orange dashed curves are extrapolations to larger depths.
}
\vspace{1cm}
\label{fig_power_peakminusbackground_depth_all_mvals}
\end{figure*}

Figure~\ref{fig_power_peakminusbackground_depth_all_mvals} shows the signal power (Eq.~\ref{eq_power_normalized_lorentz}). The quantity $\mathcal{P}_{\text{signal}}$ typically decreases from the surface toward a depth of $\SI{8}{\mega\metre}$, significantly as shown by the errors. Even further inside the Sun the power often increases again and reaches near-surface or even higher values. The 1$\sigma$ errors shown in this plot give the standard deviation, but they do not indicate $\SI{68.3}{\percent}$ probability intervals, since the power distribution is non-Gaussian (power cannot be negative). More information about error estimation can be found in App.~\ref{app_error_estimation_validation}.

\citet{Provost1981} presented a theoretical argument that the Rossby wave eigenfunctions for the horizontal displacement are proportional to $r^m$ under the assumption that the modes are incompressible. Thus, under this theory, the radial vorticity is expected to be proportional to $r^{m - 1}$. To compare this to our observations, we perform a fit of the form $const. \times r^{2\alpha}$ to $\mathcal{P}_{\text{signal}}$ within the dashed black lines ($0.7$ to $\SI{7.4}{\mega\metre}$) where the RDA is more reliable (see Fig.~\ref{fig_rotation_systematics}). We assume that the data points are uncorrelated in depth. Obviously, the fit does not reproduce the increase of power at large depths. 

\begin{figure}
\centering
\includegraphics[width=\hsize]{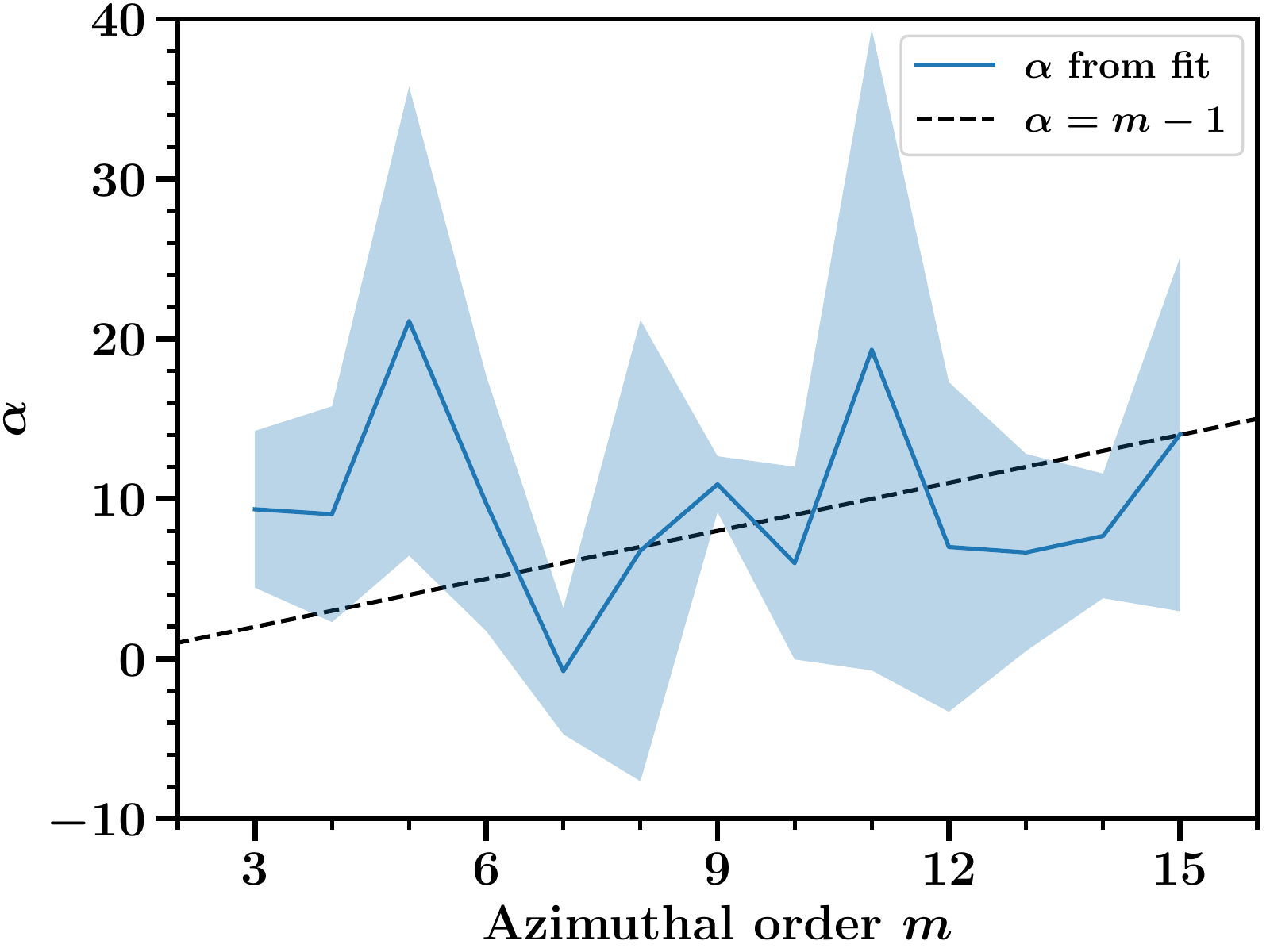}
\caption{
Exponent $\alpha$ as a function of $m$, measured in the top $\SI{7.4}{\mega\metre}$ (blue line) and 1$\sigma$ error (blue shaded area). The dashed black line corresponds to the model $\alpha = m - 1$, obtained under the assumption of non-divergent motions \citep{Provost1981}.
}
\label{fig_power_slope_fit}
\end{figure}

Figure~\ref{fig_power_slope_fit} compares the observed and theoretical exponent $\alpha$. The fitted exponent has very large error bars. It is consistent with the theoretical model from \citet{Provost1981}, but also with the absence of any trend with $m$. Although the exponent depends strongly on the fit range because of the kink at roughly $\SI{5}{\mega\metre}$ in Fig.~\ref{fig_power_peakminusbackground_depth_all_mvals}, we also do not find inconsistency with a flat dependence on $m$ within other fit intervals. Thus the current error estimates do not allow a definitive statement on the radial dependence of Rossby waves.

\section{Summary}
\label{sect_summary}

We build on \citetalias{Loeptien2018}, who investigated Rossby waves mostly using granulation tracking, by studying several Rossby wave properties via the analysis of radial vorticities computed from RDA at different depths and LCT at the surface. We obtained several new results: independently the latitudinal eigenfunctions with RDA (including a more complete, complex-valued description of the eigenfunctions), and the Rossby wave power and phase depth dependence.

We calculated latitudinal eigenfunctions of Rossby waves from the radial vorticity maps via the covariance between the equator and different latitudes and from the singular vectors of an SVD. We confirmed the shape of the real part of the eigenfunction from \citetalias{Loeptien2018}, who used the covariance method on symmetrized LCT data. We also saw consistency between covariance and SVD results, except for $m = 5$ and $m = 6$, where the SVD eigenfunctions had maxima around $\pm 10$-$\SI{15}{\degree}$ instead of at the equator. The shape of the real part of the latitudinal eigenfunctions seen for most $m$ indicates that the Rossby waves have maximum amplitudes near the equator, as found by \citetalias{Loeptien2018}. The imaginary part appears to be mostly positive for low $m$ ($3 \leq m \leq 6$); this part varies around zero for intermediate $m$ ($7 \leq m \leq 9$) and is mostly negative for high $m$ ($10 \leq m \leq 15$). A nonzero imaginary part may be due to attenuation and to the interaction of the waves with large-scale flows. In particular, the interaction of viscous Rossby waves with latitudinal differential rotation leads to the formation of critical layers at intermediate latitudes (Gizon \& Fournier, priv. comm.).

We defined and measured characteristic parameters for the real part of the eigenfunctions and we found that the width at an eigenfunction value of $0.5$ (the HWHM) decreased with $m$, in contrast to the $m$-independent width at a value of $0$ and the latitude and value of the eigenfunction minimum. We also decomposed the eigenfunctions into associated Legendre polynomials and saw that the real part is dominated by $\ell = m$ and $\ell = m + 2$ contributions, while the imaginary part consists mostly of $\ell = m + 4$ and $\ell = m + 2$ contributions for low and high $m$, respectively.

We compared rotation rates from ring-diagram data and global modes as functions of depth and saw a small offset at small depths and disagreement at large depths, but most importantly inconsistency of different ring-diagram longitudes. This indicated systematic effects in the ring-diagram pipeline (see also \citealt{Komm2015}).

We studied the Rossby wave power and phase depth dependence in detail for the first time. The phase at the peak is stable with depth, in contrast to the phase of the background. The background power almost monotonically increases with depth, while the signal power decreases toward a depth of $8$-$\SI{9}{\mega\metre}$ and then increases again. The radial eigenfunctions of the Rossby waves are (at small depths) consistent with a power-law decrease, in particular both with the theoretical \citet{Provost1981} model (exponent $m - 1$) and an $m$-independent exponent. However, the \citet{Provost1981} model is based on assumptions that are not exactly correct for the Sun (e.g., uniform rotation). We can constrain the radial dependence of the eigenfunctions only very weakly owing to the high uncertainties on the observed exponents.

The analysis presented in this paper implicitly makes the assumption that the Rossby wave eigenfunctions are separable in depth and latitude. Our data show a similar latitude dependence for the different depths, separability thus appears to be a reasonable assumption. The results shown in this work motivate further research on Rossby wave eigenfunctions, which is a necessary condition for the interpretation of the measured mode frequencies.

\begin{acknowledgements}
B.~Proxauf is a member of the International Max Planck Research School for Solar System Science at the University of G{\"o}ttingen; he conducted the data analysis and contributed to the interpretation of the results and to the writing of the manuscript. We thank \mbox{Z.-C.}~Liang for providing help with the fitting of the modes in frequency space and V.~B{\"o}ning for providing beta values for App.~\ref{app_rda_inversion_issues}. The HMI data are courtesy of NASA/SDO and the HMI Science Team. The data were processed at the German Data Center for SDO funded by the German Aerospace Center DLR. We acknowledge partial support from the European Research Council Synergy Grant WHOLE SUN \#810218.
\end{acknowledgements}

\bibliographystyle{aa}
\bibliography{literature}{}

\begin{appendix}

\section{Issues of the ring-diagram inversions}
\label{app_rda_inversion_issues}

In this Appendix, we discuss two issues regarding the ring-diagram pipeline inversions. In order to obtain local velocities at a certain measurement depth $r$, the reported pipeline velocities must be multiplied by $r/R_\odot$. Additionally the pipeline inversion does not take the quantity $\beta_{n \ell}$ (see, e.g., \citet{Aerts2010}, Eq.~3.357) into account and thus the reported inversion velocities $u_x$ are slightly incorrect.

To see this, we study a simple case. For now, let us assume that the ring diagrams are not tracked. The frequency perturbation $\delta \omega_{n \ell m}$ of the mode indexed by radial order $n$, angular degree $\ell$, and azimuthal order $m$ due to a radial differential rotation rate $\Omega(r)$ is
\begin{equation}
\delta \omega_{n \ell m} = m \beta_{n \ell} \int_0^{R_\odot} K_{n \ell}(r) \Omega(r) dr,
\label{eq_freq_splitting_rotation}
\end{equation}
where $K_{n \ell}$ is the normalized rotation kernel for that mode, i.e., $\int_0^{R_\odot} K_{n \ell}(r) dr = 1$ (see, e.g., \citet{Aerts2010}, Eq.~3.358). On the other hand, ring diagrams assume that the velocity mode fits $U_{x,\,n \ell}$ are equal to a radial integral over the true velocity flow field $u_x(r)$ weighted by flow sensitivity kernels. Based on inspection of the pipeline, we think that the used HMI kernels are normalized rotation kernels $K_{n \ell}$ from Eq.~\ref{eq_freq_splitting_rotation}. Thus
\begin{equation}
U_{x,\,n \ell} = \int_0^{R_\odot} K_{n \ell}(r) u_x(r) dr.
\label{eq_ring_fits_kernel}
\end{equation}
To connect the two equations in a simple case, consider the Doppler shift of a sectoral ($\ell = m$) mode as seen by a ring diagram at the equator, i.e.,
\begin{equation}
U_{x,\,n \ell} k_x = \delta \omega_{n \ell m} = m \beta_{n \ell} \int_0^{R_\odot} K_{n \ell}(r) \Omega(r) dr.
\end{equation}
In this equation, $k_x$ is the wavenumber in the prograde direction, which is related to $m$ via $k_x = m/R_\odot$. We conclude that
\begin{equation}
U_{x,\,n \ell} = \beta_{n \ell} \int_0^{R_\odot} K_{n\ell}(r) R_\odot \Omega(r) dr.
\label{eq_ring_fits_kernel_beta}
\end{equation}
This is not consistent with Eq.~\ref{eq_ring_fits_kernel} since $\beta_{n \ell}$ is missing from Eq.~\ref{eq_ring_fits_kernel}. Additionally we see that $u_x(r)$ should be interpreted as $R_\odot \Omega(r)$ and not as the local linear velocity $r \Omega(r)$.

To see what happens if the tracking rate is not zero, we now suppose that we track at rotation rate $\Omega_{T}$. Equation~\ref{eq_ring_fits_kernel_beta} then becomes
\begin{equation}
\tilde{U}_{x,\,n \ell} = \left[ \beta_{n \ell} \int_0^{R_\odot} K_{n \ell}(r) R_\odot \Omega(r) dr \right] - R_\odot \Omega_{T},
\end{equation}
where $\tilde{U}_{x,\,n \ell}$ is the ring measurement in the rotating frame. We now define the local deviation from the tracking rate $\delta \Omega(r) = \Omega(r) - \Omega_{T}$. Then we obtain
\begin{equation}
\tilde{U}_{x,\,n \ell} = \left[ \beta_{n \ell} \int_0^{R_\odot} K_{n \ell}(r) R_\odot \delta \Omega(r) dr \right] + (\beta_{n \ell} - 1) R_\odot \Omega_{T}.
\label{eq_ring_fits_kernel_beta_tracking}
\end{equation}
The first term is the same form as in Eq.~\ref{eq_ring_fits_kernel_beta}, while the second term is an offset that depends on $n$ and $\ell$.

The conversion factor $r/R_\odot$ is multiplied onto the data before any analysis is performed for this paper; see Sect.~\ref{sect_data_methods_rda_overview}. The offset due to $\beta_{n \ell}$ depends on the set of mode ring fits, but it should be mostly time-independent, since the ring-diagram mode set does not vary much with time. Thus the time-dependent Rossby waves should not be sensitive to this effect and the only affected result in this paper should be the comparison of rotation rates in Fig.~\ref{fig_rotation_systematics}.

To estimate the effect of $\beta_{n \ell}$ on the inversion result, for a given input flow $u_x$ we generate artificial ring fits $\tilde{U}_{x,\,n \ell}$ via Eq.~\ref{eq_ring_fits_kernel_beta_tracking}, on which we run the ring-diagram inversion module to retrieve the output velocities. To compute $\tilde{U}_{x,\,n \ell}$, we assume a depth-independent flow equivalent to the tracking rate (sidereal Carrington rate), thus $\delta \Omega(r) = 0$. We thus check only the second term of Eq.~\ref{eq_ring_fits_kernel_beta_tracking} and neglect that $\beta_{n \ell}$ also appears in the first term as a scaling factor. However, the effect due to the second term should be much larger than that due to the first term, as $\Omega_T$ is much larger than $\delta\Omega(r)$ for the ring diagrams.

We use $\beta_{n \ell}$ values provided by V.~B{\"o}ning (priv. comm.). These were computed from eigenfunctions obtained from the Aarhus adiabatic oscillation package (ADIPLS; \citealt{Christensen-Dalsgaard2008,Christensen-Dalsgaard2011}). We lose roughly $\SI{25}{\percent}$ of the original ring-fit modes, as we only have $\beta_{n \ell}$ values up to frequencies of $\SI{5}{\milli\hertz}$. However, this does not critically change the mode set used during the inversion. We replace the actual pipeline ring fits with the artificial data. We leave all other data, including uncertainties on mode-fit velocities, as is and perform the inversion. The aforementioned conversion factor of $r/R_\odot$ is multiplied onto the output velocities $u_x$.

We see that the effect of $\beta_{n \ell}$ does not depend much on depth and that the retrieved $u_x$ are on the order of only $\SI{1.5}{\metre\per\second}$ (equivalent to roughly $\SI{0.1}{\percent}$ of the tracking rate). The reason for this is that the inversion gives much more weight to the high $\ell$ modes for which the uncertainties are comparatively small. These modes typically have $\beta_{n \ell}$ values around $0.999$, thus $1 - \beta_{n \ell} \sim \SI{0.1}{\percent}$. We performed this check exemplarily for a ring-diagram tile at the first time step in our dataset (May, 20, 2010) at the point ($\lambda = \SI{0}{\degree}, \varphi = \SI{0}{\degree}$). However, tests using different tiles show that this result does not depend much on time or disk position.

The effect of the pipeline inversion not accounting for $\beta_{n \ell}$ is thus an underestimation of the true velocity fields by roughly $1$-$\SI{2}{\metre\per\second}$, or equivalently approximately $\SI{0.4}{\nano\hertz}$. This difference would be visible in Fig.~\ref{fig_rotation_systematics}, but does not change our main conclusions.

\section{Interpolation and apodization of ring-diagram velocities}
\label{app_rda_processing_steps}

We interpolate the ring-diagram velocities separately in time and longitude (see Sect.~\ref{sect_data_methods}) with different functions, depending on the number of available data points:
\begin{alignat*}{3}
&\text{-- $\geq 4$ data points:} \quad &&\text{cubic splines} \\
&\text{-- $3$ data points:} \quad &&\text{quadratic splines} \\
&\text{-- $2$ data points:} \quad &&\text{linear splines}
\end{alignat*}
Before we interpolate the ring-diagram velocities to the surface equatorial rotation rate, we apodize these velocities with a raised cosine in angular great-circle distance $\rho$ to the point ($\lambda = \SI{0}{\degree}, \varphi = \SI{0}{\degree}$), see Sect.~\ref{sect_data_methods}), as follows:
\begin{equation}
H(\rho) =
\begin{cases}
1 & \text{if } |\rho| \leq \frac{1 - \beta}{2T}, \\
\frac{1}{2} \left\{ 1 + \cos \left[ \frac{\pi T}{\beta} \left( |\rho| - \frac{1 - \beta}{2T} \right) \right] \right\} & \text{if } \frac{1 - \beta}{2T} < |\rho| \leq \frac{1 + \beta}{2T}, \\
0 & \text{else},
\end{cases}
\end{equation}
where $\beta$ defines the steepness of the raised cosine flanks. We choose $\beta = 0.3$. The quantity $T$ defines the central position of the flanks. We choose $T$ such that zero is reached at $\rho = \SI{67.5}{\degree}$ (where there are no more valid pixels). Apodizing the ring-diagram velocities (with different $\beta$), or not, gives consistent results.

\section{Error estimation and error validation}
\label{app_error_estimation_validation}

\subsection{Error estimation via chunked data}

The uncertainties on all results are derived by dividing the time series of vorticity maps (in total $2448$ time steps, i.e., $102$ rotations, for RDA) into equal-size chunks and calculating the scatter over the results for the chunks. We find that for chunks longer than a few months (roughly six rotations), the Rossby wave signature is visible in the power spectra. We make a compromise between noise level and chunk statistics and divide the dataset into five chunks that are $480$ time-steps long each ($20$ rotations).

For the latitudinal eigenfunctions, where we only study the shorter LCT period (i.e., $78$ rotations), we first used four chunks of length $19$ rotations (rotation-averaged maps) and $470$ time steps (full time resolution, for RDA), but we obtained very large errors for the SVD method for different single $m$, where single chunks gave singular vectors different from the usual eigenfunction shape. We think that the noise in our filtered maps might have been detected as the dominant term in the decomposition. We thus use a chunking with three chunks of length $26$ rotations and $625$ time steps, where all chunks have the expected first singular vectors.

\subsection{Error validation via Monte Carlo simulation}

We validate the chunking approach for the depth dependence via a Monte Carlo simulation. As a plausible physical model for the Rossby wave power spectrum, we assume a Lorentzian profile and a background (constant in frequency), each with a $\chi^2$-distributed random variable (stochastic excitation). In analogy to Eq.~\ref{eq_power_fit_lorentz} we generate $1000$ realizations of synthetic data for the Fourier transform of the radial vorticity $\mathcal{F}_{\text{syn}}$ (not the power $P$), at each $m$, as
\begin{equation}
\label{eq_monte_carlo_model}
\mathcal{F}_{\text{syn},\,m}(\nu,r) = \sqrt{\frac{A_m(r)}{4(\nu - \nu_{0,\,m})^2/\gamma_m^2 + 1}} \mathcal{N}_{A,\,m}(\nu) + \sqrt{B_m(r)} \mathcal{N}_{B,\,m}(\nu,r).
\end{equation}
We fix the amplitude $A_m(r)$, background $B_m(r)$, central frequency $\nu_{0,\,m}$, and full width at half maximum $\gamma_m$ via the fit parameters for the observations. Furthermore we assume that the random variable $\mathcal{N}_{A,\,m}(\nu)$ is constant with depth, i.e., the signal is fully correlated in depth, while for the background, we take a random variable $\mathcal{N}_{B,\,m}(\nu,r)$ that is uncorrelated in depth. The random variables are complex Gaussian variables (with independent real and imaginary parts) with zero mean and unit variance, independent for each frequency. We analyze the realizations in the same way as the observations.

We observe inconsistency between two Monte Carlo estimates (by roughly $\SI{30}{\percent}$): one from chunking (like the observations), averaged over the realizations, and the other from the scatter of the power over the realizations. We find that the discrepancy is due to the temporal correlation of the different chunks; the temporal correlation matrix has values around $0.1$ on the first off-diagonal. The two quantities agree when using a weighted average with weights based on the temporal covariance matrix.

Both Monte Carlo estimates disagree with the error for the observed data. This could be due to the depth correlation of the observations. We determine a (noisy) estimate of the depth covariance and correlation matrix of the observed background power from the different background frequencies and find strong correlations between different depths, even between the largest depth and the surface, with values above $0.25$ in the off-diagonal corners. We use the observed depth covariance matrix for the Fourier transform (not the power) to correlate the Gaussian background random variable of our realizations, via a Cholesky decomposition.

Correlating the background in depth noticeably improves the agreement between Monte Carlo and observed errors. However, there is still a remaining discrepancy that can be attributed to our model, which is missing some features of the observed power spectrum. We do not account for the window function nor a background decreasing with frequency present in the observations. A detailed analysis goes beyond the scope of this paper.

\section{Relation of covariance to linear fit}
\label{app_latitudinal_eigenfunction_methods}

The covariance method (Sect.~\ref{sect_latfunc_cov}) is conceptually equivalent to a linear fit of the vorticity at each depth $r$ and latitude $\lambda$, i.e.,
\begin{equation}
\tilde{\zeta}_m^{'}(t,r,\lambda) = a_m(r,\lambda) f_m(t).
\label{eq_eigenfunc_cov_linfit}
\end{equation}
Let us for simplicity assume that the vorticity $\tilde{\zeta}_m^{'}(t,r,\lambda)$ is real. The latitude and depth dependence in this vorticity separation ansatz is contained in the fit parameter $a_m(r,\lambda)$. Let us assume that only $\tilde{\zeta}^{'}$ is uncertain. For zero-mean quantities (such as our vorticity maps $\tilde{\zeta}^{'}$), the slope of a linear fit without intercept, i.e., $a_m(r,\lambda)$, is given as
\begin{equation}
a_m(r,\lambda) = \frac{\langle \tilde{\zeta}_m^{'}(t,r,\lambda) f_m^*(t) \rangle_t}{\langle \vert f_m(t) \vert^2 \rangle_t}.
\end{equation}
Assuming the time dependence is given by the surface equatorial vorticity time series, i.e., $f_m(t) = \tilde{\zeta}_m^{'}(t,r = R, \lambda = \SI{0}{\degree})$, in Eq.~\ref{eq_eigenfunc_cov_covnorm} we can identify $a_m(r,\lambda)$ with $C_m(r,\lambda)$. Equation~\ref{eq_eigenfunc_cov_linfit} implies that $a_m(r = R, \lambda = \SI{0}{\degree})$ is unity.

The main disadvantage of the covariance method is the assumption of a noise-free vorticity at the equator, $\tilde{\zeta}_m^{'}(t,r = R, \lambda = \SI{0}{\degree})$, required so that the time-dependence $f_m(t)$ is noise-free and the vorticity $\tilde{\zeta}_m^{'}(t,r,\lambda)$ is the only uncertain quantity of the fit.

\end{appendix}

\end{document}